\documentclass{aastex631}

\begin{document}

\title{\boldmath \texttt{$\mu$-GLANCE}: A Novel Technique to Detect Chromatically and Achromatically Lensed Gravitational Wave Signals}

\author[0009-0004-4937-4633]{Aniruddha Chakraborty}
\affiliation{Tata Institute of Fundamental Research, 
Homi Bhabha Road, Navy Nagar, Colaba, 
Mumbai 400005, India}

\author[0000-0002-3373-5236]{Suvodip Mukherjee}
\affiliation{Tata Institute of Fundamental Research, 
Homi Bhabha Road, Navy Nagar, Colaba, 
Mumbai 400005, India}



\begin{abstract}

Gravitational microlensing in the wave-optics (WO) regime occurs when the Schwarzschild radius of a lensing object is comparable to or smaller than the wavelength of incoming gravitational waves (GWs), producing chromatic amplitude and phase modulations. In contrary, geometric optics effects happen when wavelength is much smaller than the lensing object leading to frequency independent amplifications and phase shifts. GWs can undergo both effects of lensing due to interactions with objects of different scales. To detect and characterize the wave-optics features from a lensed GW, we have developed a novel method \texttt{$\mu$-GLANCE} (Micro-Gravitational Lensing Authenticator using Non-modelled Cross-correlation Exploration). In this technique, we calculate the cross-correlation between the residuals from different detectors. We assign a false alarm rate to each potential WO microlensed candidate from a statistical viewpoint, depending on how many times the noise cross-correlation matches the strength of the residual cross-correlation of the candidate over a certain period of time. We show that for an event with a matched-filtering signal-to-noise ratio (SNR) close to thirty, a residual due to wave-optics lensing with an amplitude about $10\%$ of magnification $\mu\approx 3.2$ will start to show deviation from noise distribution at more than $68\%$ Confidence interval with the LIGO-Virgo-KAGRA sensitivity for the fourth observation run. This method provides the first technique to detect geometric optics and wave-optics effects from a WO microlensed GW without assuming any specific lensing model, and its application on the current and future GW data can identify events with both chromatic and achromatic lensed scenarios.
\end{abstract}



\section{Introduction}
Gravitational waves (GWs) are propagating fluctuations in the spacetime metric caused by the time-varying quadrupole moment of the mass distribution of a system. With the help of Laser Interferometer Gravitational-wave Observatories (LIGOs) present at Livingston and Hanford sites, we have confirmed the existence of GWs on September 14, 2015 \citep{LIGOScientific:2016aoc}. Until the third observational run (O3) in 2021, the LIGO-Virgo-KAGRA detector network (LVK) has more than 90 confirmed detections of GWs \citep{LIGOScientific:2018mvr, LIGOScientific:2020ibl, KAGRA:2021vkt, KAGRA:2013rdx, VIRGO:2014yos, LIGOScientific:2014pky,  Buikema_2020, KAGRA:2020tym}. These observations are caused by some of the most catastrophic processes in the universe, such as mergers of compact object binaries. These detections have opened up a plethora of explorable physics regimes, ranging from cosmology (e.g. \citep{LIGOScientific:2021aug}), compact binary population inferences (e.g. \citep{KAGRA:2021duu}) to probing fundamental physics (e.g. \citep{LIGOScientific:2021sio}). 

Gravitational lensing is the phenomenon where the presence of a massive object dictates the light-like trajectories passing close-by \citep{1992grlebookS, PhysRevLett.77.2875, PhysRevLett.80.1138, Bartelmann_2010, PhysRevD.95.044011}. The massive object bends the path of the incoming GWs, deflecting them from their original trajectories. This causes the waves to interfere, producing amplitude and phase modulation on the outgoing GW waves. Quite analogous to the different regimes of optics with electromagnetic waves, gravitational lensing of GWs can be treated in two different scenarios: (i) Geometric optics : if the Schwarzschild radius ($R^s_{lens} = \frac{2GM}{c^2}$) of a massive object (called lens hereafter) is much larger than the wavelength of the emitted GWs ($\lambda_{GW}$), i.e. $R^s_{lens} \gg \lambda_{GW}$, we can consider geometric-optics approximation. We observe a system of multiple images with GWs, with frequency-independent magnification and phase-shifts. (ii) Wave optics (WO) : if the Schwarzschild radius of the lens is of comparable size or smaller than the wavelength of the GW $R^s_{lens} \leq \lambda_{GW}$, interfering signals produce GWs with frequency-dependent amplitude modulation and phase modulation factors.

Gravitational lensing provides a unique way to infer the density profile, shape and dimension of the massive lensing objects. While electromagnetic waves lensing happens in the geometric-optics regime (due to their short wavelengths compared to the large astrophysical objects in the universe), chances of observing wave optics effects on them are negligibly small. On the other hand, GWs having larger wavelengths (minimum $\approx 10^2$ km to maximum $\approx 10^{14}$ km \citep{Bailes:2021tot}), WO microlensing of GWs is feasible. Therefore, inference of the shape, size and mass distribution of dark matter structures can be performed by studying the wave-optics lensing signatures on GWs. Using lensing of GWs in the observable frequency spectra across all current and future GW detectors, we can probe properties of massive bodies of $10^4 M_{\odot}$ to $10^{14} M_{\odot}$ mass \citep{WAMBSGANSS2006567, Bayer:2023nwm}.
Lensing of GWs offers unique probe constraints on the astrophysical population of high redshift compact objects \cite{Mukherjee:2020tvr, 10.1093/mnras/stab1980}, cosmological parameters \citep{Sereno:2011ty, Balaudo:2022znx,Jana:2022shb} to study the dark matter distribution \citep{Massey:2010hh, Basak:2021ten}, testing General Theory of Relativity \citep{Fan:2016swi, Mukherjee:2019wcg, Mukherjee:2019wfw} and to detect intermediate and primordial black holes \citep{PhysRevD.98.083005, Oguri:2020ldf}.

With the current LIGO-Virgo-KAGRA(LVK) sensitivities \citep{Abbott:2016xvh, KAGRA:2013rdx, aLIGO:2020wna}, observation of a lensed GW is very rare, having the probability of occurrence of a few parts in a thousand (around 0.1 - 0.6\%) \citep{10.1093/mnras/sty411, 10.1093/mnras/stab1980, Diego_2021}. The lensing search by the LVK collaboration on the O3 data led to no significant evidence of any lensing of GWs \citep{2023arXiv230408393T} using existing methods \citep{10.1093/mnras/stab1991, Wright:2021cbn, PhysRevD.107.123015, Hannuksela:2019kle, Dai:2020tpj, Singer:2019vjs}. A follow-up study on the O3 events also did not found any significant evidence of the observation of a lensing event so far \citep{10.1093/mnras/stad2909}. With improved detector sensitivities in the upcoming O5 run and with around $\approx 1000$ GW events, we expect to witness at least one lensed GW.

In the cold dark matter model, we  expect large dark matter halos to have smaller sub-haloes inside it \citep{Zavala:2019gpq}. So while the halo overall as a lensing object produces geometric-optics effects on the GWs, the sub-haloes (or any present substructures) can contribute to WO microlensing modifications \citep{Fairbairn:2022xln, Diego:2019lcd, PhysRevD.101.123512, Meena:2023qdq}. Such scenarios not only allow us to study the properties of the dark matter (DM) halo itself but also let us understand the substructures present within the DM halo. However, the WO microlensing modifications can be an order of magnitude smaller than the geometric-optics regime \citep{Takahashi:2003ix}, therefore making it very hard to be observed. However, with advancements in the GW detectors and with the inclusion of next-generation GW observatories, we expect to be able to distinguish such minuscule modifications on the waveform. Such lensing modifications can impose false bias on estimation of the GW source intrinsic parameters \citep{Mishra:2023ddt}.

In this perspective, we have developed \texttt{$\mu$-GLANCE}, Micro-Gravitational Lensing Authenticator using Non-modelled Cross-correlation Exploration, a technique to detect WO microlensed GW signals. The technique essentially relies on the residual cross-correlation among different detectors. Performing cross-correlation on the residuals, helps us understand the common unmodelled features present in the signals as observed in all detectors. To summarize, we perform two different Bayesian parameter estimations on the data: one with the WO microlensing hypothesis and the other with no lensing hypothesis. Therefore, in the strong presence of the WO microlensing in GW signal (reflected by the support of the WO microlensing parameters in the probability distributions), we can detect them straightforwardly. However, if there is no strong support for the WO microlensing parameters, we shift our focus on the source parameters obtained using the no WO microlensing hypothesis and construct the best-fit signal. We cross-correlate between the residuals at different detectors and assign a false alarm rate (FAR) associated. If the FAR is less than the threshold FAR, set at $10^{-3}$ per yr, we call the event as an WO microlensed candidate. On the other side, if there are multiple images of the same GW i.e. in the geometric-optics scenario, we apply \texttt{GLANCE} \citep{Chakraborty:2024net} to detect lensed GW signals. We perform cross-correlation on the reconstructed polarizations and its deviation from the noise cross-correlation helps us determine the significance of the event pair being actually lensed.

Recently, there has emerged several search pipelines to detect WO microlensed GWs. WO microlensing, if present in the data, would have some changed characteristics in the spectrogram of the GW. Therefore, deep learning based methods to detect WO microlensing signatures from the spectrogram has been performed \citep{Kim:2022lex}. WO microlensing induces a bias to the source parameter estimation. Therefore, to detect WO microlensing, a joint parameter estimation based technique has also been implemented to obtain the source and lensing template parameters together \citep{Wright:2021cbn}. Also in the presence of a macrolens and microlens substructures inside, geometric-optics and WO effects can be both present on a GW signal. This would enhance the detection of WO microlensing due to multiple images of the WO microlensed GW signal; extracting parameters from those multiple signals would allow to better constrain the WO microlensing parameters \citep{Seo:2021psp}. 

In this work, we have presented \texttt{$\mu$-GLANCE} a novel technique to detect chromatically and achromatically lensed GWs. This work is organised as follows:  in section \ref{sec:2}, we discuss gravitational lensing of GWs and its geometric-optics and WO microlensing regime. Here, we also motivate about the particular lensing amplification model used in this work. We discuss the salient features of 
\texttt{$\mu$-GLANCE}, its strengths and weaknesses in detecting lensed GWs in section \ref{sec:3}. In section \ref{sec:4} we discuss the methodology used and the mathematical framework behind the cross-correlation based technique. We apply the technique on simulated GW data in section \ref{sec:5} and show its performance of detecting WO microlensed signals. In section \ref{sec:6}, we present the effect of variation of the source and lens parameters and how that supports us for a confident WO microlensing detection. To remove the inference biases on the source parameters caused by WO microlensing, we jointly estimate the source and lensing template parameters in section \ref{sec:jointpe}. The false alarm rate associated with a WO microlensing detection is calculated in section \ref{sec:8}. In section \ref{sec:9}, we discuss the possible caveats and the future prospects of \texttt{$\mu$-GLANCE} in the detection of a WO microlensed GW.

\section{Salient aspects of \texorpdfstring{\texttt{$\mu$-GLANCE}}{mu-GLANCE}
}\label{sec:2}

\begin{figure}
    \centering
    \includegraphics[width=0.8\linewidth]{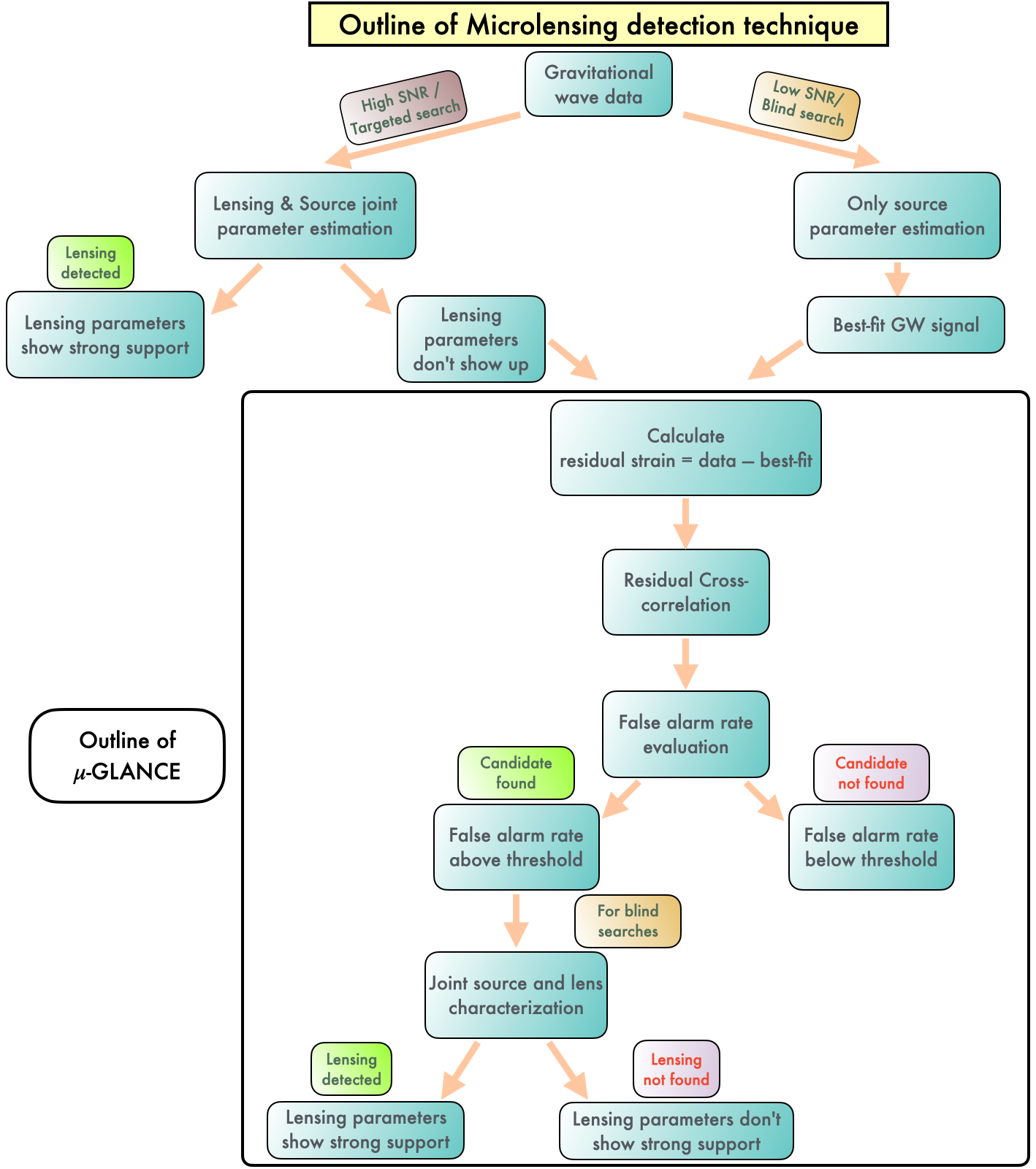}
    \caption{The figure shows the outline for WO microlensing searches of GWs using \texttt{$\mu$-GLANCE}. For targeted WO microlensing searches on GW events (The targetted search is performed on a GW event, whose the sky localization is of the order of a few sq. degrees upto tens of sq. degrees consisting of lensed galaxies of \textit{relevant} redshifts), we first perform the joint source and lensing template parameter estimation (PE) on the data, for the events with high matched-filtering SNR. If we observe strong support for the lensing parameters, we can directly claim the event to be a potential WO microlensing candidate. If no strong support is shown towards the lensing parameters, it is tested though residual cross-correlation in $\mu-$\texttt{GLANCE}. The technique also incorporates any low matched-filtering SNR events to perform a blind WO microlensing search from the data. It performs cross-correlation on the residuals from different detectors. The residuals are constructed by subtracting the best-fit strain (in the unlensed hypothesis) from the data. After the cross-correlation strength has been compared against the noise cross-correlation though its false alarm rate (FAR), a significance of the event for its WO microlensing candidature is evaluated. If the event qualifies strong support though the residual cross-correlation, it is declared as WO microlensing candidate. If the joint parameter estimation has not been performed already, we perform it to constrain the lensing effects on the waveform along with the characterization of the source. A confident detection is approved only when the lensing parameters show deviation from the null lensing hypothesis regime.}
    \label{fig:0}
\end{figure}

\texttt{$\mu$-GLANCE} can look for vast scenarios of WO microlensing signatures present in the data through residual cross-correlation. It can both detect and characterize the lensing signal from the GW data. We show a flowchart describing various parts of \texttt{$\mu$-GLANCE} in figure \ref{fig:0} and describe some of the salient aspects of the method below: 
\begin{enumerate}
    \item \textbf{Model-independent nature of the algorithm:} Any detection method requires to be model-independent or support a large variety of models. Given the limited class of only spherically symmetric lenses studied so far, we cannot make a model-dependent claim of WO microlensing detection. That is why \texttt{$\mu$-GLANCE} follows a first-in-class model-independent approach for WO microlensing search from the data. Through cross-correlation between the residuals from different detectors, \texttt{$\mu$-GLANCE} performs cross-correlation to look for the common non-localized WO microlensing features buried in the data.
    
    \item \textbf{Removal of noise artifacts though cross-correlation:} The detection method needs to be very robust, and it should suppress noise artifacts. Cross-correlation-based \texttt{$\mu$-GLANCE} searches for the presence of any underlying similarities/patterns even at low strengths, while uncorrelated noise from the different detectors is suppressed heavily. Here, we compare the strength of the event cross-correlation as compared to the background noise fluctuations. This helps us understand the significance of an event. Any non-Gaussianity present in the data e.g., glitches which are not correlated among the two detectors, thus \texttt{$\mu$-GLANCE} can handle these situations well.
    
    \item \textbf{Allowance for different detector pairs for a stronger claim on detection:}The detection should have the strength of incorporating multiple detectors for a stronger detection claim. Cross-correlation picks up any similar feature present between the residual pair from a pair of detectors. \texttt{$\mu$-GLANCE} combines all combinations of the pair of detectors and makes a stronger claim for detection if a common feature is observed in different pairs of detectors.
    
    \item \textbf{Minimization of waveform systematics:}  Residual morphology depends on the chosen waveform model used for the best-fit subtraction from data. As a result, waveform-based systematics can pose a potential threat to WO microlensing detection. This is why we have made \texttt{$\mu$-GLANCE} in a way that can minimize waveform-based artifacts in the residual cross-correlation. Thus, to check for waveform systematics, \texttt{$\mu$-GLANCE} can perform cross-correlation on residuals subtracting best-fit templates from different waveform models. It helps to benchmark our confidence with the results coming from a specific model and how much it can vary by the changing waveform model. On top of that, by cautious selection of only the part of the waveform from an f$_{\rm min}$ to f$_{\rm max}$ in the range over which the waveform systematics are limited. This is built in \texttt{$\mu$-GLANCE} so that it only considers this well-modelled part of the waveform, which is least affected by waveform systematics \footnote{We would like to mention the necessity of very accurate waveform models to capture all features of the GW. For our current analysis, use \texttt{IMRPhenomXPHM} \citep{Pratten:2020ceb} waveform model that incorporates eccentric, precessing black hole binary and calculates the waveform for higher modes starting from $(l,m) = (2,2) $. We have considered the  [(2, 2), (2, 1), (3, 3), (3, 2), (4, 4)] and the corresponding negative $m$-modes for the generation of the GW-signal. The inclusion of higher order modes, though marginal in GW data analysis with LVK, would have a significant impact in the next detectors such as Cosmic Explorer \citep{Evans:2021gyd} and Einstein Telescope \citep{Maggiore:2019uih}.
}. 
    
    \item \textbf{Joint inference of source properties and lensing template properties - to remove WO microlensing biases:} WO microlensing modulations are frequency-dependent, often mimicking the effects of GW source properties such as spin, eccentricity, etc. Therefore, it can apply bias to the inference of the source properties. Therefore, to remove lensing biases, \texttt{$\mu$-GLANCE} performs a joint exploration of source properties and lensing template parameters. We construct a very generic model that captures the WO microlensing amplification effects caused by an astrophysical object on the GW. Therefore, in comparison to geometric-optics lensing searches as in \texttt{GLANCE} \citep{Chakraborty:2024net}, the WO microlensing search is not completely unmodelled. The WO microlensing parameters show up in the joint source and lensing template parameters' distributions when there are WO microlensing modulations present at significant strength. If the modulations are weakly present, the joint source and lensing template parameters distribution may not be able to constrain the WO microlensing parameters well, and in such cases, we solely rely on the residual cross-correlation signal to detect any WO microlensing-like signatures.
\end{enumerate} 

These salient aspects of this method make it useful to detect lensing signals from different lensing scenarios, along with mitigating noise contaminations. By this technique, the presence of both WO and geometric-optics lensing signature can be measured from the data, along with distinguishing from unlensed events with a false alarm rate. A detailed discussion on this is presented in Sec. \ref{sec:8}. For the scenarios of only geometric-optics lensing and without WO effects (multiple-image system), the performance of this technique is shown by \citet{Chakraborty:2024net}.  

\subsection{Comparison with the existing WO microlensing-based searches:}

This new method \texttt{$\mu$-GLANCE} estimates the best-fit unlensed waveform to obtain the residuals from each detector for a GW event. These residuals are then studied through cross-correlation for any common signature not captured in the waveform model. The relative strength of the cross-correlation signal as compared to the noise fluctuations helps us quantify the detection significance. This \texttt{$\mu$-GLANCE} is the very first model-independent search technique to look for WO microlensing signatures. From the entire third run of the LVK observatories, the LVK collaboration searched for potential WO microlensed signals from the data \citep{LIGOScientific:2021izm, LIGOScientific:2023bwz}. It estimated the Bayes factor, which shows the support for the lensing hypothesis as compared to the non-lensing hypothesis by comparing the respective likelihoods. The work observes no strong evidence for any WO microlensed candidate by observing the Bayes factors. However, the work has considered point-mass lensing objects for their analysis. The point-mass lens is not a good approximation for its spherical symmetry, a very uncommon trait for an astronomical object. 

Recently, machine-learning-based techniques \citep{Kim:2022lex} have been applied to the data to search for point-mass lens effects on the data. These methods use spectrogram (q-transformed data, which maps the evolution of the GW event through the time-frequency bins). However, such searches are not conclusive since the mass distribution of lensing objects is most of the time not spherically symmetric. Also, the waveform model for the training dataset is \texttt{IMRPhenomPv2}, which is not inclusive of all such different models. 

Joint estimations of the source and the lensing template parameters are also performed for WO microlensing detection in a previous work \citep{Wright:2021cbn}. The work incorporates the wave-optics effects of the GW for point-mass, singular-isothermal-sphere (SIS), and Navarro-Frenk-White (NFW) mass distribution profiles. However, these distributions are all spherically symmetric, for objects of sizes of the GW wavelength, the assumption of the spherical symmetry of the lens cannot be confirmed, Therefore, a more rigorous way would either require sophisticated lens profiles or need a completely model-free approach, which is not followed in this technique.

\section{Basics of Gravitational Lensing: in Wave-optics and Geometric-optics lensing Regimes}\label{sec:3}

In the presence of matter, the propagation of GW depends on the gravitational potential of object(s) along its trajectory. The equation describing the spatial variation of the GW amplitude is given by \citep{1992grlebookS},
\begin{eqnarray}
  \left(\nabla^2 + \frac{\omega^2}{c^2} \right) \tilde{\phi} (\vec{r}, \omega) = \frac{4\omega^2 U(\vec{r})}{c^4} \tilde{\phi}(\vec{r}, \omega),
\end{eqnarray}
where $\nabla^2$ is the 3-dimensional Laplacian operator, $\omega$ is the angular frequency ($=2\pi f$) of the GW, $U(\vec{r})$ is the gravitational potential (assumed not to be varying in the scale of the passing time of the GW) of the lens and the GW strain in the frequency domain is given by, $h_{\mu \nu } (\omega, \vec{r}) = e_{\mu \nu} \tilde{\phi}(\omega, \vec{r})$ where $ \tilde{\phi}(\omega, \vec{r})$ consists of amplitude and phase of the GW and $e_{\mu \nu}$ is the associated polarization.

\begin{figure}
    \centering
    \includegraphics[width=0.7\linewidth]{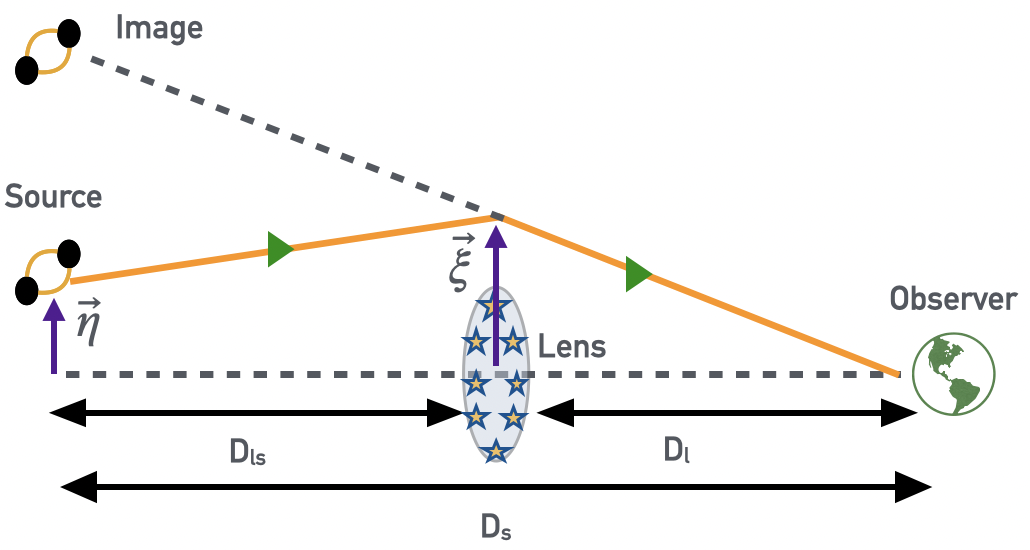}
    \caption{The figure shows a schematic diagram for the gravitational lensing with a thin lensing approximation. The thin lens approximation assumes the deflection of the ray happens only at the lens plane.}
    \label{fig:deflection}
\end{figure}

The thin lens approximation assumes that the deflection of the GW caused by the lens happens at the lens plane, a plane passing through the centre of the lens at which the lensing object in projected. In figure \ref{fig:deflection}, we have showed a ray diagram schematic for gravitational lensing. Let, $D_l$ be the angular diameter distance \footnote{Angular diameter distance is a distances inferred from the angular size of the sources. See \citep{Hogg:1999ad} } between the observer and the lens, $D_s$ is the angular diameter distance between the observer and the source and $D_{ls}$ is the angular diameter distance between the lens and the source. $\vec{\eta}$ is the position of the source in the source plane. The source plane origin is located at the intersection the extended straight line joining the the observer to the centre of the lens. $\vec{\xi}$ is the impact parameter of the incoming ray in the lens plane. The origin of the lens plane is located at the centre of the lens. We have defined two dimensionless vectors $\vec{x}$ and $\vec{y}$ as following:
\begin{eqnarray}
    \vec{x} = \frac{\vec{\xi} } {\xi_0} \quad \text{and} \quad \vec{y} = \frac{ D_{l} \vec{\eta}}{\xi_0 D_s},
\end{eqnarray}
where $\xi_0$ sets the characteristic distance scale of the lens system.

The thin lens approximation is valid when the dimension of the lens is much smaller than the total travelled distance by the GW. The depth of gravitational potential which in turn controls the strength of the amplifications leads to the classification of gravitational lensing as strong lensing or microlensing. The amplification factor is defined as $A(f) = \frac{\tilde{\phi}^{\rm lensed}(f, \vec{r})}{\tilde{\phi}^{\rm unlensed} (f, \vec{r})}$. 
The term $\tilde{\phi}^{\rm lensed} (f, \vec{r})$ at the numerator, is dependent on the relative time delay between the arrival of GWs through different spatial trajectories. This time delay includes the geometric time delay between GWs travelling from different deflected trajectories and the Shapiro time delay, caused by slowing of time at higher spacetime curvature regions. The total time delay is given by \citep{1992grlebookS, Takahashi:2003ix},
\begin{equation}
    t_d(\vec{x}, \vec{y})=\dfrac{D_s \xi_0^2}{c D_l D_{ls}}\left(1+z_l\right)\left[\frac{1}{2}|\vec{x}-\vec{y}|^2-\psi(\vec{x})+\Phi_m(\vec{y})\right] \quad,
    \label{eq:td}
\end{equation}
where $z_l$ is the redshift of the lens corresponding the angular diameter distance of the lens $D_l$ , $c$ be the speed of light in vacuum. The first term provides the geometric time-delay and the second term provides the Shapiro time-delay where $\psi(\vec{x})$ provides the gravitational potential of the lens in the lens plane. The third term with $\Phi_m(\vec{y})$ is optional, to set the minimum $t_d$ value to zero. In presence of gravitational lensing, GW waves are time-delayed and deflected therefore the addition of GWs in different phases causes interference of lensed GW in the wave-optics limit. 

\begin{figure}
    \centering
    \includegraphics[width=0.8\linewidth]{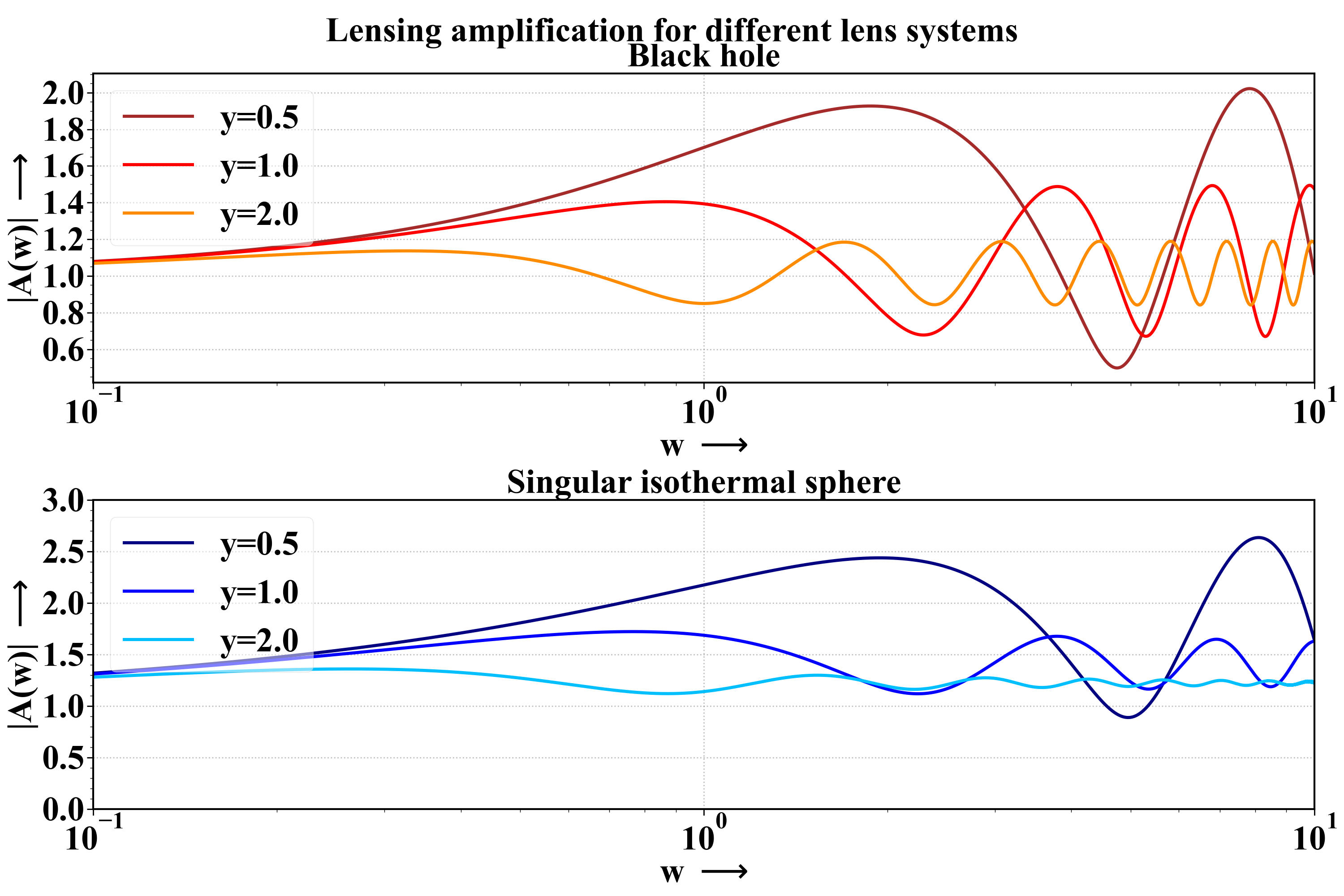}
    \caption{The figure shows the modulus of the amplification factor for point-mass lens and SIS lens for three different values of y: y=0.5, y=1.0 and y=2.0, in the range $w \in [0.1, 10]$}
    \label{fig:amp}
\end{figure}

For EM wave passing through an aperture, the amplification factor of the wave at the other side is obtained by using the Kirchhoff integral. A similar expression can be applied for GW lensing. Using the time-delay expression, we can find the amplification factor as, 
\begin{equation}
    A(f)=\frac{D_s \xi_0^2}{c D_l D_{ls}} \frac{f}{i} \int d^2 \vec{x} \exp \left[2 \pi i f t_d(\vec{x}, \vec{y})\right].
\end{equation}
Accounting for the expansion of the universe, the frequencies $f$'s are all redshifted by a factor of $(1+z_l)$. For a spherically symmetric lens potential, projected on the lens plane, the expression of the amplification factor can be written as, 
\begin{equation}
    A(w)=-i w e^{\frac{i w y^2}{ 2}} \int_0^{\infty} dx \left[x J_0(w x y) e^{ \left[i w\left(\frac{1}{2} x^2-\psi(x)+\Phi_m(y)\right)\right]} \right],
\end{equation} 
here $J_0$ is the spherical Bessel function of order zero, and $w$ is the dimensionless frequency defined as, $w= \dfrac{ 8 \pi G M_{lz} f}{c^3}$ where $M_{lz} = M_l (1+z_l)$ is the redshifted mass of the lens, $f$ be the frequency of the GW and G is the gravitational constant.

Different lensing regimes can also be described by the dimensionless parameter $w$. When $w \leq 1$, we are in the WO microlensing regime, the amplification factor amplitude $|A(w)|$ and its phase $\theta_A (w) = - i \rm{log_e} \left(\frac{A(w)}{\lvert A(w) \rvert} \right)$ are very oscillatory in nature. However, in the limit $w \gg 1$, we are in the geometric-optics regime. Both $|A(w)|$ and $\theta_A (w)$ converge to a single value independent of the frequency of the GWs. The amplification factor in the geometric-optics lensing regime (for any generic mass or size of lens) becomes, 
\begin{equation}
A(f)=\sum_j\left|\mu_j\right|^{1 / 2} \exp \left[2 \pi i f t_{d, j}-i \pi n_j\right],
\end{equation}
where, the flux-magnification of the j-th image is $\mu_j = 1 / \operatorname{det}\left(\frac{\partial \vec{y}}{\partial \vec{x}_j}\right)$, and  $t_{d,j} = t_d (\vec{x_j}, \vec{y})$ and $n = 0, 1/2, 1$ for minimum, saddle point, maximum of the $t_d(\vec{x}, \vec{y})$ function. They are known as type-I, type-II and type-III images respectively. Transforming the amplification back in the Fourier-conjugate i.e. time-domain, we can write the lensed wave amplitude in the time-domain is then represented as, 
\begin{figure}
    \centering
    \includegraphics[width=0.8\linewidth]{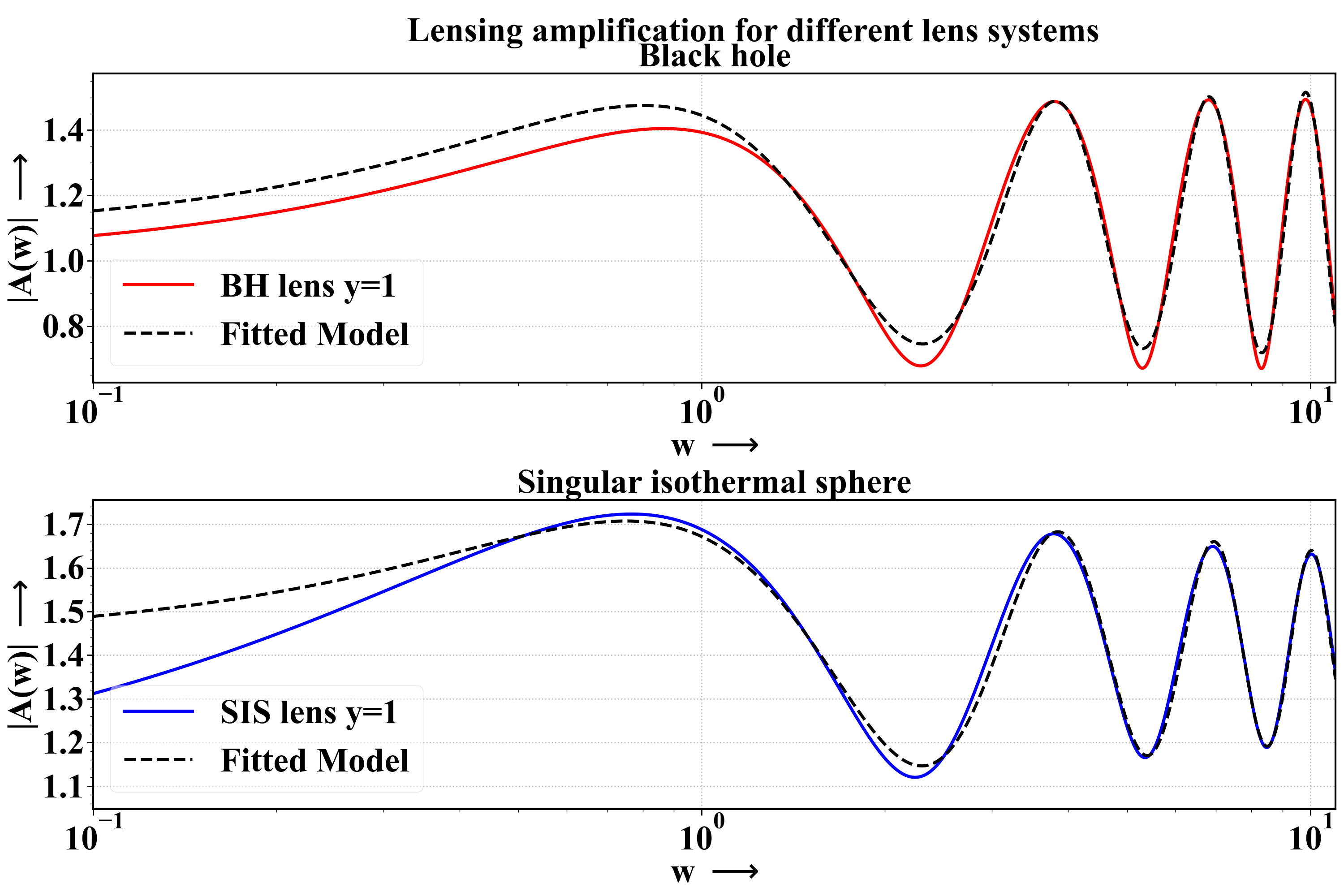}
    \caption{The figure shows the capability of the amplification model in capturing the necessary features present in the WO microlensing amplification. We picked up two cases from the figure \ref{fig:amp} and fitted it with our generic amplification model. The fitting parameters are mentioned in the text.}
    \label{fig:amp_fit}
\end{figure}
\begin{equation}
\phi^{ \rm lensed} (t, \vec{r})=\sum_j\left|\mu_j\right|^{1 / 2} \phi^ {\rm unlensed}\left(t-t_{d, j}, \vec{r}\right) \exp \left[-i \pi n_j\right]
\end{equation}

For the point mass lens and SIS lens, we have analytic solutions \citep{PhysRevD.108.043527}. We plot the solutions as a function of $w$ with $y$ as a parameter in figure \ref{fig:amp}. However, such analytical solution may not exist for more general lens including asymmetric shape and non-spherical mass distribution. Also, the impact of baryons on the lens profile, and consequently on the WO microlensing signature, remains largely uncertain. As a result, for lensing search pipeline to discover signal with wave-optics effects, it is important for it to capture any general lensing scenario. We note that any wave-optics effect is captured by a scale close to unity and constant periodicity in the frequency-domain (both captured by the misalignment parameter $y$ and the dimensionless frequency $w \propto M_{lz}f$). Also these wave effects decreases as $w$ becomes much larger than unity, this is the geometric-optics regime. Driven by these observations we follow a lens-model agnostic approach, and construct a frequency modulated template that captures most (if not all) of the features of WO microlensing of GWs. The model is described as:
\begin{equation}
A(f) = a\left[1+b e^{-kf} \cos\left(\frac{2 \pi f}{f_0} + \phi \right)\right], 
\end{equation}
where $a$ is lensing amplitude \footnote{If geometric-optics (GO) lensing present together with WO microlensing, it would also contribute to this amplitude, $ a = a_{GO} \times a_{WO} \equiv \sqrt{\mu} \times a_{WO}$}, $b$ captures perturbation scale at which the WO microlensing in the amplification varies, $f_0$ denotes characteristic frequency of the oscillations in amplification, $\phi$ denotes the phase shift due to lensing, and $e^{-kf}$ ensures that the amplification slowly converges to $a$ as $f$ is much greater than $f_0$. In the later sections, we use this amplification model to understand the different aspects of a WO microlensing detection and characterization. In figure \ref{fig:amp_fit}, we have demonstrated the capability to contain all the features in a WO microlensing amplification for the case with $y=1$ with a point-mass lens and an SIS lens (With lens of mass $500 \rm M_{\odot}$ at a redshift of $z_l = 0.2$) . The fitting parameters for the point-mass amplification profile are $(a,b, k , f_0, \phi) =(1.113, 0.322, -0.0008, 40.383\, \rm{Hz}, 4.613)$ and the fitting parameters for the SIS amplification profile are $(a,b, k , f_0, \phi) =( 1.421, 0.206, 0.002, 41.565\, \rm{Hz}, 4.745)$ \footnote{Considering no geometric-optics lensing $\sqrt{\mu}=1$, these fitting parameters have been obtained.}. We observe that, the amplification model with appropriate fitting parameters can nicely mimic the WO microlensing amplification\footnote{In the current setup, the WO microlensing model does not capture the phase modulation. But it can be trivially included by a few additional variables to capture the phase modulations, in expense of computational additional time. Thus to demonstrate this technique, we have taken a model with amplitude modulation.}.

\section{Methodology}\label{sec:4}
To obtain the residuals, first
we perform a Bayesian parameter estimation technique to constraint the parameters of interest (masses, spins, inclination, distance, RA, Dec, coalescence time, polarization angle) with the unlensed hypothesis. We use the values of the medians of the distributions to obtain a best-fit strain $h_{BFi}(t)$ at detector $i$ from the data. We subtract this from the data, called $d_i(t)$, to obtain the residual for the detector $i$, denoted as $R_i(t)$. Therefore the residual is defined as, 
\begin{equation}
    R_i(t) \equiv d_i(t) - h_{BF i}(t) = r_i(t) + n_i(t),
\end{equation}
here $r_i(t)$ denotes the residual signal and $n_i(t)$ denotes the noise present in the data. 
We define the cross-correlation between two residual data as from two detectors denoted by $\{x, x'\}$,
\begin{align}
    D_{x x'} (t) &\equiv R_x \otimes R_{x'} = \frac{1}{\tau} \int_{t-\tau/2} ^{t+\tau/2} r_{x} (t') r_{x'} (t' + \Delta t_{\rm det}) dt', \nonumber\\
    &= S_{x x'}(t) + N_{x x'}(t) + P_{x x'}(t) + Q_{x x'}(t), \nonumber\\
    & \approx S_{x x'}(t) + N_{x x'}(t), 
\end{align}
where $\tau $ is the timescale over which the cross-correlation is performed and $\Delta t_{\rm det}$ be the time-delay between the arrival of the signal (thus the residual) between the detectors. The second expression contains four terms are : $S_{xx'} \equiv r_x \otimes r_{x'}$, $N_{xx'} \equiv n_{x} \otimes n_{x'}$, $ P_{xx'} \equiv r_x \otimes n_{x'}$, $ Q_{xx'} \equiv n_{x} \otimes r_{x'}$ respectively, where `$\otimes$' denotes the cross-correlation between them. The final expression is approximated when the cross-correlation timescale ($\tau$) is sufficiently large, comparable to the duration of the signal, the cross-terms do not contribute much.

Therefore, if the absence of any WO microlensing, the best-fit would be able to capture the GW waveform, making the cross-correlation signal $S_{xx'}$ negligible. However, if WO microlensing present in the GW signal, all its features are not captures by the best-fit signal, therefore it shows up in the cross-correlation. In this analysis, we have ignored waveform systematics and its impact on lensing. The cross-correlation technique can find any non-general relativistic unmodelled signatures present in the signal as well \citep{Dideron:2022tap}. 

A similar but distinct detection approach has been taken with strong lensing regime when there are multiple images of a GW signal separated by time delays from seconds to years. We have showed in our previous work \texttt{GLANCE} \citep{Chakraborty:2024net} that, such strongly lensed GW signals in the geometric-optics limit can be detected by performing cross-correlation between the reconstructed one-polarization signals of the images using a detector pair at two different times. We defined lensing SNR to quantify how much the cross-correlation signal differs from the noise cross-correlation values. The method demonstrates that for type-I and type-II pair (or type-II and type-III pair) of images, the cross-correlation shows up by among the different polarizations, otherwise it shows up in the same polarization. A joint parameter estimation was performed to explore the degeneracies between the source properties and the lensing effects and thus helping in removing the lensing bias. Therefore to look for the geometric-optics lensing aspects from the data, we have polarization cross-correlation based technique \texttt{GLANCE} and to look at the WO microlensing aspects, we have now developed residual cross-correlation based technique \texttt{$\mu$-GLANCE}.

\section{Application of \texorpdfstring{\texttt{$\mu$-GLANCE}}{mu-GLANCE} on simulated GW data}\label{sec:5}

\begin{figure}
    \centering
    \includegraphics[width=0.85\linewidth]{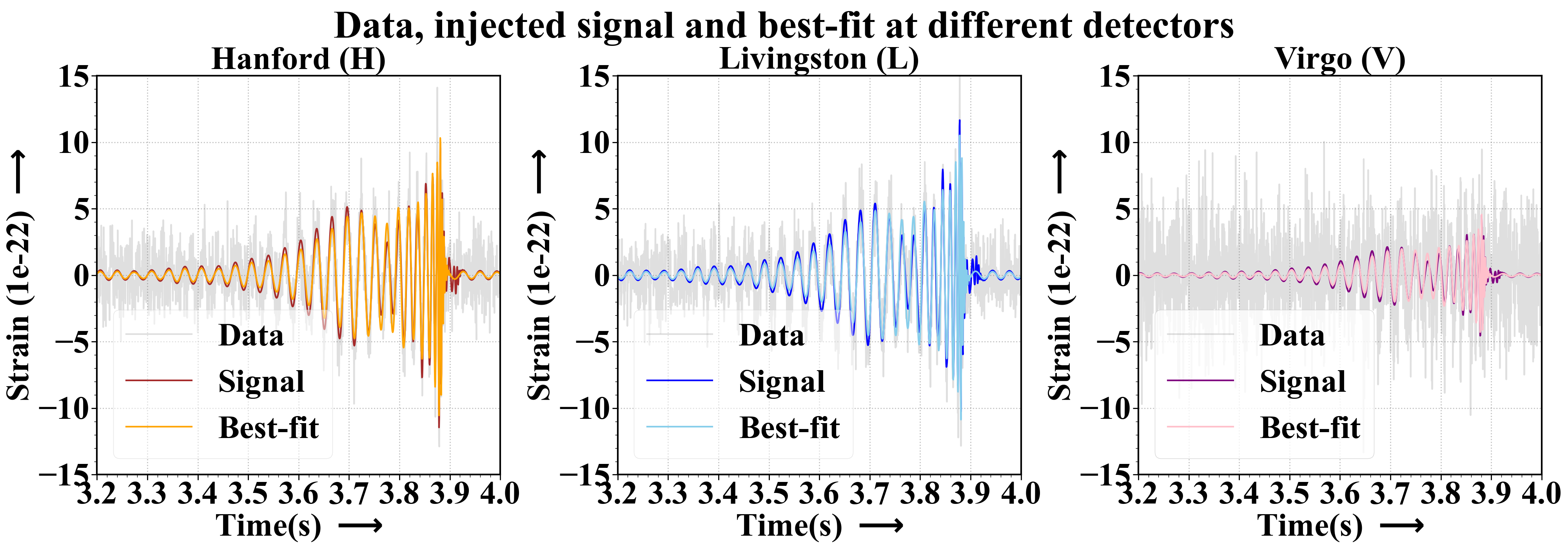}
    \caption{The figure shows the data, signal injected and the best-fit signal for the LIGO-H, LIGO-L and Virgo-V observatories. The best-fit is obtained first by performing an estimation of the source parameters (masses, aligned spins, luminosity distance and inclination angle) with an unlensed hypothesis and then using the medians of the posterior distributions.}
    \label{fig:1}
\end{figure}

As a demonstration of the technique, we generated GW data using a system of binary black holes with masses $35 M_{\odot}$ and $25 M_{\odot}$ at $1$ Gpc \footnote{1 Gpc = $10^9$ parsec (pc), where 1 pc = 3.086e+16 metre} luminosity distance with inclination angle $0.5$ and dimensionless aligned spins of the primary and the secondary as $0.1$ and $0.2$ respectively. We generate the waveform using the phenomenological model \texttt{IMRPhenomXPHM} \citep{Pratten:2020ceb} and used the analytic noise power spectral density (PSD) model \texttt{AdVO4T1800545} and \texttt{AdvVirgo} \footnote{See \citep{advLIGO} and \citep{Manzotti:2012uw} respectively} to generate the noise for the LIGO and Virgo detectors respectively. We choose the amplification factor \footnote{A more generic approach consists of a phase $\phi$ in the cosine term of the amplification factor and estimating that parameter as well, but this form of the amplification bears all the necessary modifications required. In the appendix \ref{app:0}, we have showed that the exclusion of the phase $\phi$ does not change any significant results performed in this analysis.} as 
\begin{equation}\label{eq:11}
    A(f; a, b, k, f_0) = a\left[1+b e^{-kf} cos \left(2 \pi \frac{f}{f_0}\right) \right].
\end{equation}
The value of the strong lensing magnification $\sqrt{\mu} \equiv a $ is chosen to be $1.8$ with WO microlensing perturbation parameter $b = 0.4$, WO microlensing oscillation scale parameter $f_0 =30$ Hz and the microlensing WO to geometric-optics transition parameter $k=0.004$. We obtained the best-fit strain without the hypothesis of WO microlensing present in the data.  The data, injected signal and the best-fit (obtained in the frequency domain and then reverted back in time domain) is shown in the figure \ref{fig:1}. The contribution of the GW observatory KAGRA is not discussed in this work, since the KAGRA sensitivity is less in comparison to the currently ongoing fourth observation run sensitivity of LIGO and Virgo detectors and therefore inclusion of KAGRA will not make any significant differences in the results. 

\begin{figure}
    \centering
    \includegraphics[width=0.85\linewidth]{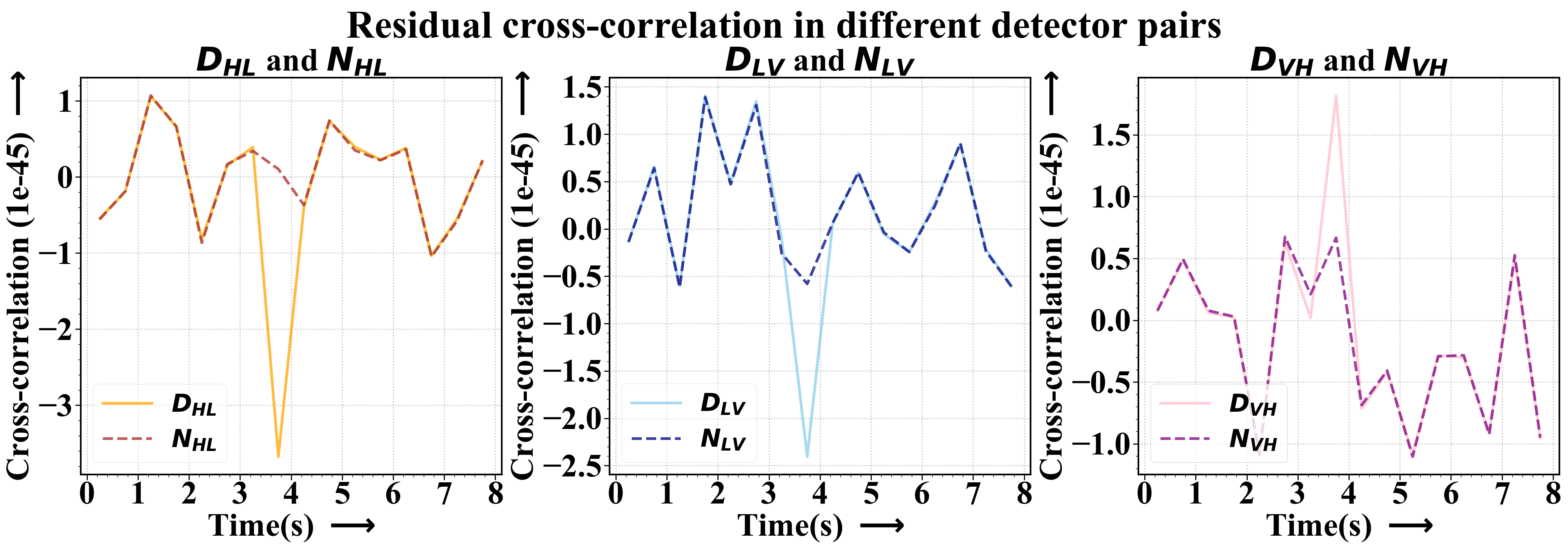}
    \caption{The figure shows the cross-correlation signal between WO microlensed GW residuals for three detector configuration using H, L, and V, when the GW signal is present ($D_{xx'}$) and when the GW signal is not present ($N_{xx'}$). The cross-correlation timescale is chosen as $\tau =0.5s$.}
    \label{fig:x}
\end{figure}

We calculate the residual at each detector by subtracting the best-fit signal from the data. We align the residuals in time by considering the RA and Dec distributions of the GW event in the sky-map. We smoothen the data by passing it through a bandpass with lower cut-off and upper cut-off frequencies respectively at $30$ Hz and $512$ Hz \footnote{The choice of the band is very specific to this example, with lower mass systems, we may need to set up the upper cut-off at higher frequencies}.  In figure \ref{fig:x}, we have shown the residual cross-correlation in the presence of the WO microlensed GW signal and in the absence of the signal, for pairs formed within LIGO-H, LIGO-L and Virgo-V. The signal is injected close to the fourth second \footnote{The merger happens at the 3.9th second of the signal.} in a $8$ second long time-series data, that is why we observe a cross-correlation peak at that position. The cross-correlation timescale $\tau = 0.5s$, since that is the typical duration of the signal. It is noteworthy to mention that, the matched-filtering signal to noise ratio (SNR) for the residual drops around 70-80\% when compared with the matched-filtering SNR obtained from the data. 

\begin{figure}
    \centering
    \includegraphics[width=0.85\linewidth]{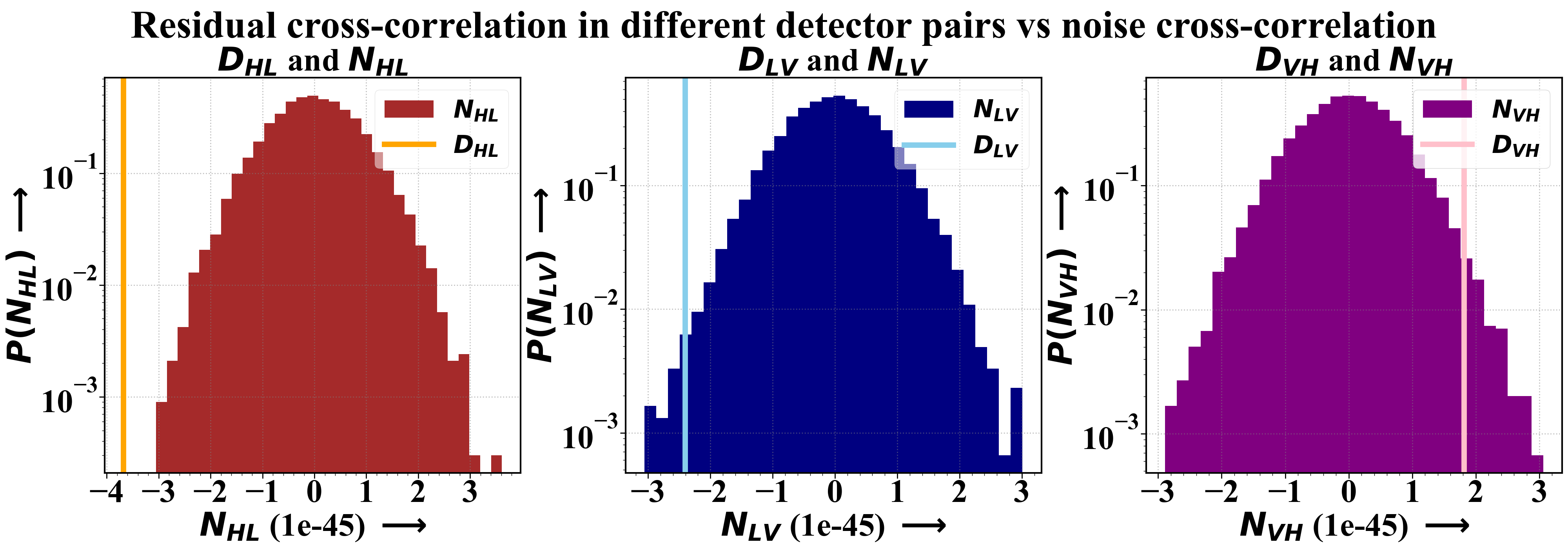}
    \caption{The figure shows the residual cross-correlation with H, L and V residual strains as opposed to the noise cross-correlation distribution. The cross-correlation timescale chosen here is $\tau =0.5s$. The choice of $\tau$ is decided by the total mass of the system, with lower $\tau$'s chosen for higher total mass systems. The lines from show the strength of the cross-correlation and the histogram is constituted from the noise cross-correlation values.}
    \label{fig:2}
\end{figure}

To understand how the residual cross-correlation peak stands out in comparison with the background noise distribution, we have plotted the residual cross-correlation at the timestamp where the signal is present as a line and showed the noise cross-correlation distribution in figure \ref{fig:2}. The noise cross-correlation histogram distribution is obtained by generating 1000 noise realizations of $16$ second each for all three detectors. Then cross-correlation is performed by taking a pair of noise realization for two different detectors and this is repeated for all the realizations in all possible different detector combinations. These noise cross-correlation values are plotted as the histogram shown in figure \ref{fig:2}. 

\begin{figure}
    \centering
    \includegraphics[width=0.8\linewidth]{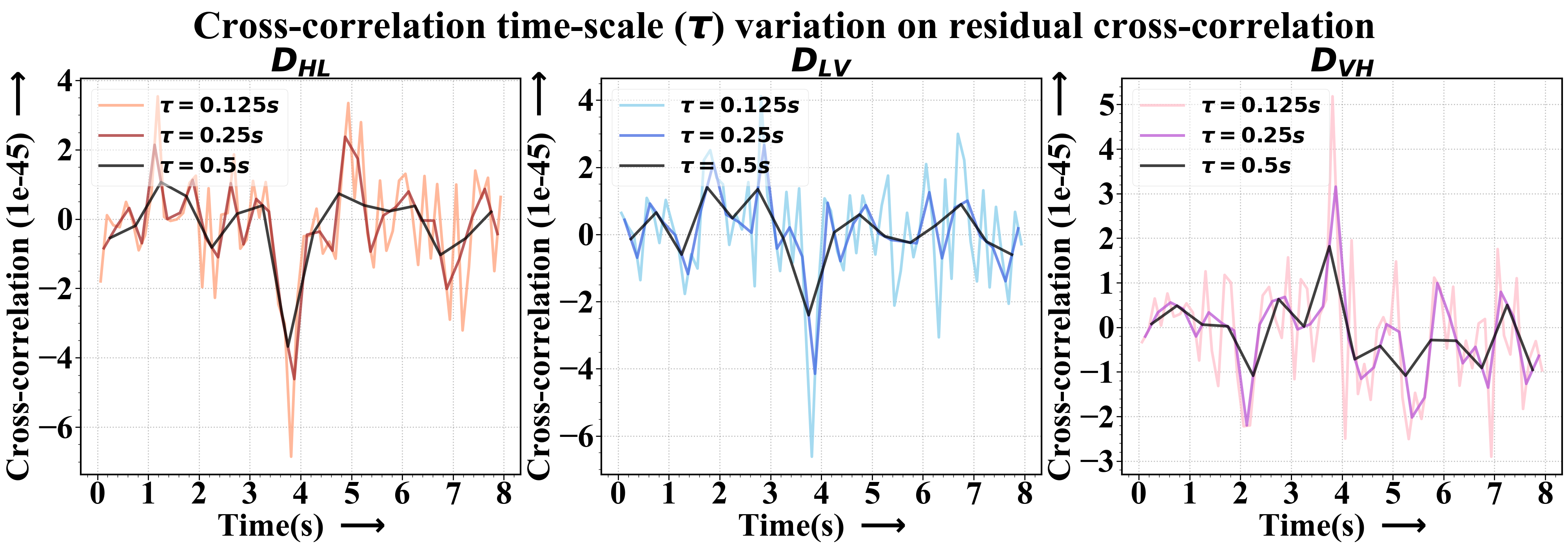}
    \caption{The figure shows the variation of the residual cross-correlation $D_{xx'}$ with cross-correlation timescale. The noise cross-correlation values follow a random distribution with a mean tending towards zero. We observe that as the cross-correlation timescale is increased, the noise-fluctuations go down, making the the residual cross-correlation more prominent.}
    \label{fig:y}
\end{figure}

Furthermore, to understand the dependency of the residual cross-correlation signal on the cross-correlation timescale, we have plotted the residual cross-correlation with varying the cross-correlation timescale parameter ($\tau$) in figure \ref{fig:y}. It can be clearly observable that as we have increased the timescale $\tau$, the random noise fluctuation go down faster than the residual cross-correlation signal, making the residual cross-correlation appear more prominent.

\section{Cross-correlation Signal Dependency on GW Source and Lens Parameters}\label{sec:6}

\begin{figure}
    \centering
    \includegraphics[width=0.85\linewidth]{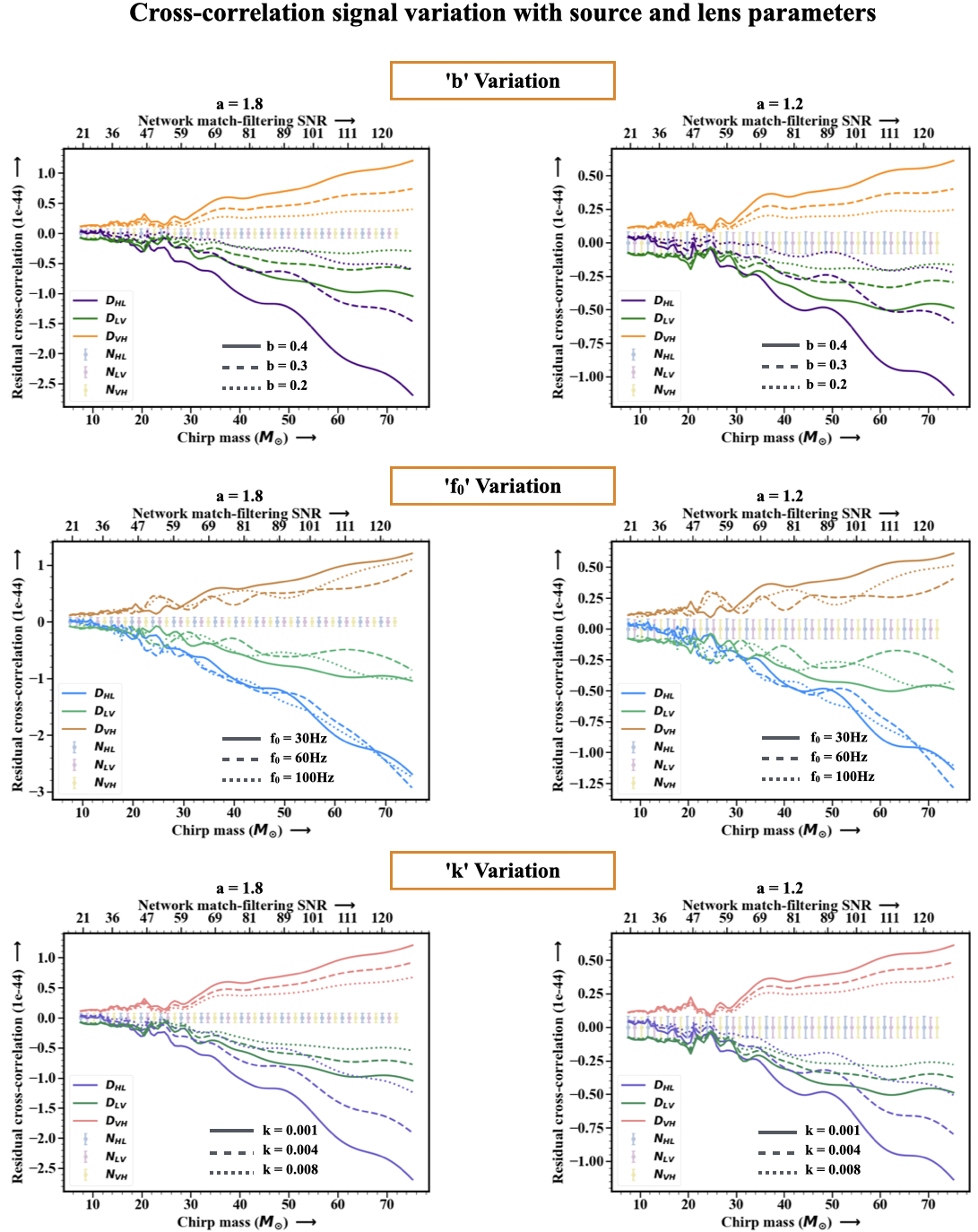}
    \caption{The figure shows the variation of the residual cross-correlation with respect to variation in the matched-filtering SNR of the GW signal by varying the chirp mass (detector frame) and lensing parameters (b, $f_0$ and k). The cross-correlation timescale is chosen to be $\tau = 0.5s$. This shows that where the residual cross-correlation deviates from the noise cross-correlations significantly (when the SNR is large and also for large b, small k), the detection becomes efficient. However, in the regime, when the deviation is small (when the SNR is small and also for small b, large k), the detection is inefficient. The effect of the $f_0$ on the results is not very decisive.}
    \label{fig:3}
\end{figure}

The residual cross-correlation $(D_{xx'})$ depends on the performance of the unlensed hypothesis parameter estimation to obtain the best-fit GW strain. The better the strain data matches with the best-fit strain is, the smaller is the residual and the residual cross-correlation. However, it is shown in the section \ref{sec:jointpe} that, GW source parameters are well constrained with a model considering no lensing hypothesis even when there is WO microlensing present. This happens because the source parameters are completely non-degenerate with the lensing template considered in the analysis. Therefore, the residual cross-correlation depends on the strength of the signal, formally given by the matched-filtering SNR, that defines how well can we obtain the best-fit. Thus since the WO microlensing parameters are not captured by the lens parameters in the unlensed hypothesis PE, there exists strong dependency of the residual cross-correlation on the values of the WO microlensing parameters. Therefore, to observe how well can we distinguish a lensing modulation, we have checked the variation of the residual cross-correlation with respect to lensing parameters. In figure \ref{fig:3}, we have shown the detectability of the residual cross-correlation with respect to the background noise cross-correlation, when no signal is present in the data. 

The shaded regions in figure \ref{fig:3}, shows the 1-$\sigma$  bounds on the noise cross-correlation distribution. The lines show the residual cross-correlation value when the signal consists of WO microlensing amplification. We varied the chirp mass to increase the strength of the signal. We fixed the value of $a$ to 1.2 and 1.8, and then varied the parameters of $b$, $f_0$ and $k$. The luminosity distance of the source is kept fixed at $1$ Gpc. The values of the WO microlensing parameters are chosen in consistency with the numbers obtained in \ref{sec:3} when we fitted the amplification models of point-mass and SIS with our template in figure \ref{fig:amp_fit}.

In the first row of the plots, we have varied the value of the WO microlensing amplitude parameter $b$, keeping $f_0$ and $k$ fixed at $30$ Hz and $0.001$ respectively. We observe that for $a=1.8$, $b=0.4$ at chirp mass as low as $20M_{\odot}$ the cross-correlation signal at H-L lies 3-$\sigma$ above, however at L-V pair the cross-correlation signal comes above the noise distribution 3-$\sigma$ at chirp mass $40M_{\odot}$, same for V-H is at around $25M_{\odot}$. As the value of b is decreased to a lower value, the minimum chirp mass at which the cross-correlation signal is well distinguishable \footnote{By "well distinguishable" we mean that, the residual cross-correlation is at least 3-$\sigma$ away from the noise cross-correlation distribution} of the BBH shifts towards higher masses. Similarly for $a=1.2$ with $b=0.4$, $f_0=30$ Hz, $k=0.001$ we observe that for H-L, L-V, and V-H the cross-correlation signal is distinguishable at $30 M_{\odot}$, $50 M_{\odot}$ and $35 M_{\odot}$ respectively, which is moved towards higher values as the value of b is decreased.

In the second row of the plots, we have varied the value of the WO microlensing oscillation parameter $f_0$, keeping $b$ and $k$ fixed at $0.4$ and $0.001$ respectively. Unlike the cross-correlation dependency on $b$, we don't observe a direct observational effect of $f_0$ variation, for any of the detector pairs. For both the cases with $a=1.8$ and $a=1.2$, we observe that the three lines corresponding to $f_0=30$ Hz, $f_0=60$ Hz, $f_0=100$ Hz, the lines are intertwined, not significantly altering the minimum chirp mass at which the cross-correlation signal is distinguishable. We understand that, the value of $b$ is more critical to determining WO microlensing in GW data as compared to $f_0$.

In the third row of the plots, we have varied the value of the WO microlensing oscillation converging parameter $k$ (as the GW frequencies move towards geometric-optics lensing regime), keeping $b$ and $f_0$ fixed at $0.4$ and $30$ Hz respectively. Quite similar the cross-correlation dependency on $b$, we observe a strong dependency of $k$ variation. For the case with $a=1.8$, we can distinguish the cross-correlation signal for H-L, L-V, and V-H at $35 M_{\odot}$, $45 M_{\odot}$ and $40 M_{\odot}$ respectively for the value of $k=0.008$. The minimum chirp mass is moved towards the lower end as the value of $k$ is decreased. Similarly, with $a=1.2$, we can distinguish the cross-correlation signal for H-L, L-V, and V-H at $40 M_{\odot}$, $60 M_{\odot}$ and $45 M_{\odot}$ respectively for the value of $k=0.008$. We understand that the effect of the transition parameter $k$ is similar to the effect of perturbation parameter $b$, therefore making it a very crucial parameter for the detection of WO microlensed signal.

\section{Joint exploration of GW Source Parameters and Lensing Template Parameters} \label{sec:jointpe}

Lensing can affect the inference of the source properties obtainable from the GW signal through existing parameter estimation techniques. Therefore, a correct approach for inferring source parameters, not just includes the source intrinsic and extrinsic parameters, rather it as well consists of the lens-induced characteristic parameters on the GW signal. We perform Bayesian parameter estimation technique in the frequency domain to jointly explore the parameter space occupied by the source and lensing template parameters. To showcase the capability of the technique, we choose a simple model for the lensing amplification factor as shown in equation \ref{eq:11}.
Here, we have kept the value of $\phi=0$ for the simulation set for the purpose of demonstrating the method. The results including $\phi$ is shown in the appendix \ref{app:0}. For future analysis of GW data, it can be trivially included in the Bayesian analysis framework. We choose the unlensed signal model given as
\begin{equation}
    h_x(f; m_1, m_2, d_{\rm source}, \iota, s_{1z}, s_{2z}) = \sum_{i= +, \times} F_x^{i}h_x^{i}(f; m_1, m_2, d_{\rm source}, \iota, s_{1z}, s_{2z}), 
\end{equation}

\begin{figure}
    \centering
    \includegraphics[width=0.95\linewidth]{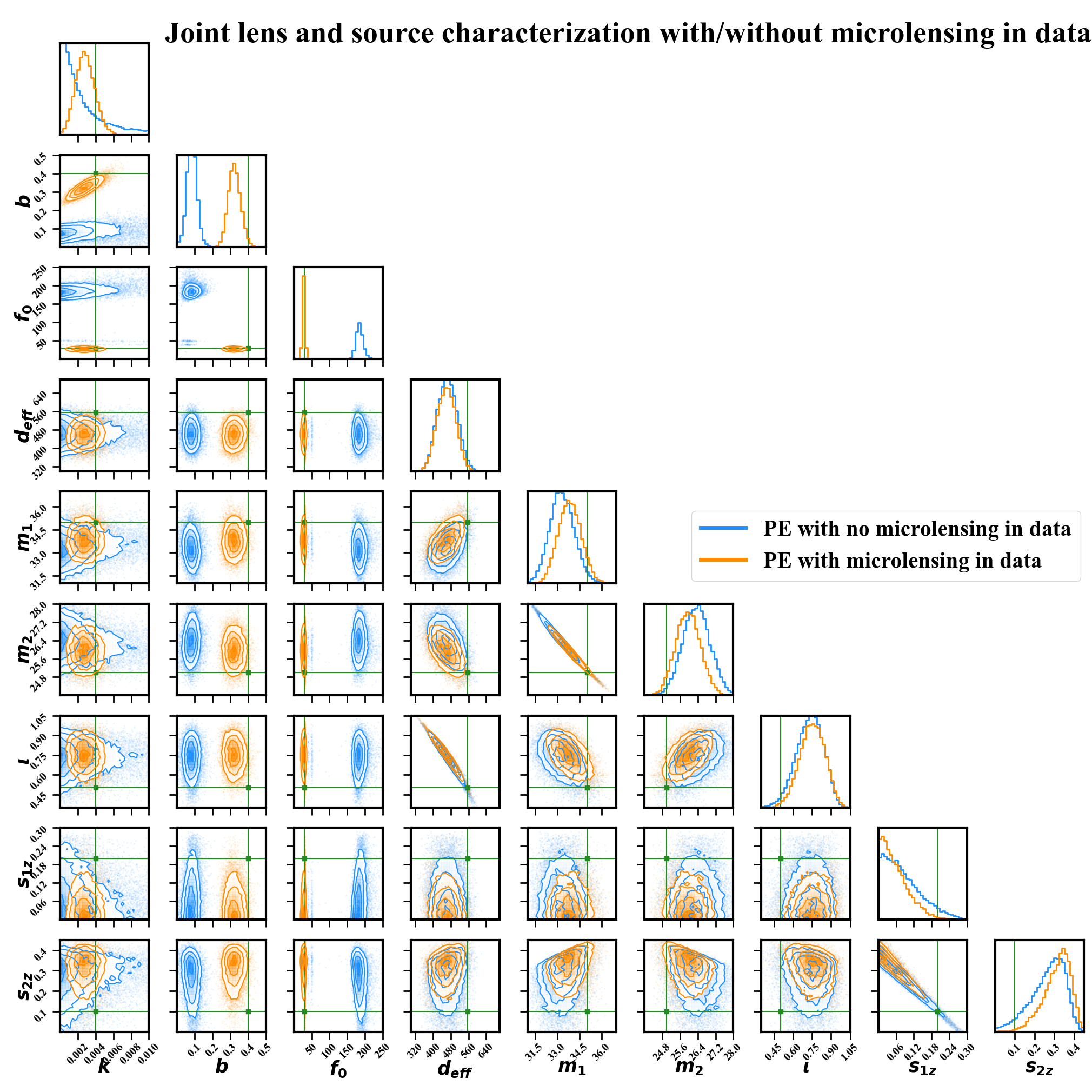}
    \caption{The corner plot shows the distribution of source and lensing template parameters obtained from a joint parameter estimation technique in presence and in absence of WO microlensing. It shows that for WO microlensing present and WO microlensing absent cases, the source parameter estimations are not very different, although quite expectedly, the lensing parameters are occupying different regions of the parameter space.}
    \label{fig:param_estim}
\end{figure}

where $i$ denotes the polarization and $x$ denotes the detector and $F^{+}$ and $F^{\times}$ are the antenna patterns associated with the plus and cross polarization. Here $m_1$, $m_2$ are the masses of the primary and the secondary of the compact binary, $d_{\rm source}$ is the luminosity distance to the source, $\iota$ is the inclination angle (the angle between the orbital angular momentum and the line of sight). $s_{1z}$ and $s_{2z}$ denote the aligned spin components of the primary and the secondary. 
The lensed waveform in the frequency domain is the product of the amplification factor and the unlensed signal. Therefore, the lensed signal model is 
\begin{equation}
\begin{aligned}
    h^{lensed}_x(f; \{\theta\}) = \sum_{i= +, \times} F_x^{i} \Big[
    A(f; a, b, k, f_0) \cdot \\
    h_x^{i}(f; m_1, m_2, d_{\rm source}, \iota, s_{1z}, s_{2z}) \Big], 
\end{aligned}
\end{equation}
here $\{\theta\} \in \{k, b, f_0, m_1, m_2, d_{\rm eff}, \iota, s_{1z}, s_{2z}\}$. Notably, we absorb the effect strong lensing magnification $a$ into the luminosity distance of the source $d_{\rm source}$ to combine an effective distance parameter $d_{\rm eff}$.  
Therefore, in the source and lensing template parameters joint characterization we include source masses, spins, effective luminosity distance and inclination angle along with WO microlensing modulations.

We formulate the Bayesian parameter estimation using the package \texttt{BILBY} \citep{Ashton:2018jfp}, a python-based module extensively used for GW data analysis purposes. We use the waveform model \texttt{IMRPhenomXPHM} for both data generation and Bayesian parameter estimation purposes. We highlight that the estimation of source and lensing template parameters is waveform model-dependent and can vary with different waveform prescriptions due to variations in underlying physical assumptions and approximations. However, this waveform systematics effect can be mitigated by choosing a time interval between which different waveform models agree with one another. This is achieved by choosing the maximum frequency cut-off as $f_{\rm isco}$, up to which the signal is at the inspiral epoch, during which the waveform systematics are limited. For the priors, we choose uniform distribution for all parameters as mentioned in table \ref{table:priors}.

\begin{table}
\centering
\begin{tabular}{|c|c|c|c|}
\hline
\textbf{Parameter} & \textbf{Distribution Type} & \textbf{Minimum} & \textbf{Maximum} \\ \hline
$k$ & Uniform & 0 & $1 \times 10^{-2}$ \\ \hline
$b$ & Uniform & 0 & 1 \\ \hline
$f_0$ & Uniform & $20$Hz  & $1024$Hz \\ \hline
$d_\text{eff}$ & Uniform & 100 & 7000 \\ \hline
$m_1$ & Uniform & 5 & 100 \\ \hline
$m_2$ & Uniform & 5 & 100 \\ \hline
$\text{inc}$ & Uniform & 0 & $\pi$ \\ \hline
$s_{1z}$ & Uniform & 0 & 1 \\ \hline
$s_{2z}$ & Uniform & 0 & 1 \\ \hline
\end{tabular}
\caption{Prior distributions for various source and WO microlensing parameters.}
\label{table:priors}
\end{table}

The likelihood is a Gaussian and it uses the frequency-domain data and model from all three detectors (in future we can include KAGRA\citep{KAGRA:2020tym} and LIGO-India \citep{LIGO_India}). The log-likelihood is given by
\begin{equation}
\log(L) = -\sum_{f} \sum_{x} \left[ \frac{(d_{x}(f) - h_{x}^{\rm lensed} (f; \{ \theta \}))^2}{2 S_x(f) } + \frac{1}{2}\log(2 \pi S_x(f)) \right] ,
\end{equation}
where $d_x(f)$ is the data and $S_x(f)$ is the noise power spectrum density in detector denoted by $x$. 
We sample the posteriors using nested sampling algorithm \texttt{DYNESTY} \citep{Speagle_2020}. The corner plot consisting of the parameters is shown in figure \ref{fig:param_estim}.

The figure \ref{fig:param_estim} shows the distribution of parameters of the source and the lensing estimated from the data (i) when there is a WO microlensed GW signal (ii) when there is no WO microlensed GW signal. The figure has many different aspects to discuss. Firstly, when the WO microlensing was present in the data, we note that all the parameters are close to well constrained around their injected values, as compared to the other case when the data did not contain a WO microlensed GW. In the scenario without WO microlensing, $k$ value inclined towards zero, $b$ also has a median less than 0.1 and $f_0$ has peaked around 190 Hz which is higher than $f_{\rm isco}$ for these binaries and close to the maximum frequency of emission from this system. The source parameters are well constrained showing almost no deviation in the parameter distributions with WO microlensing hypothesis. The joint parameter distribution clearly indicates that there is no degeneracy among the source and lensing template parameters.

\begin{figure}
    \centering
    \includegraphics[width=0.9\linewidth]{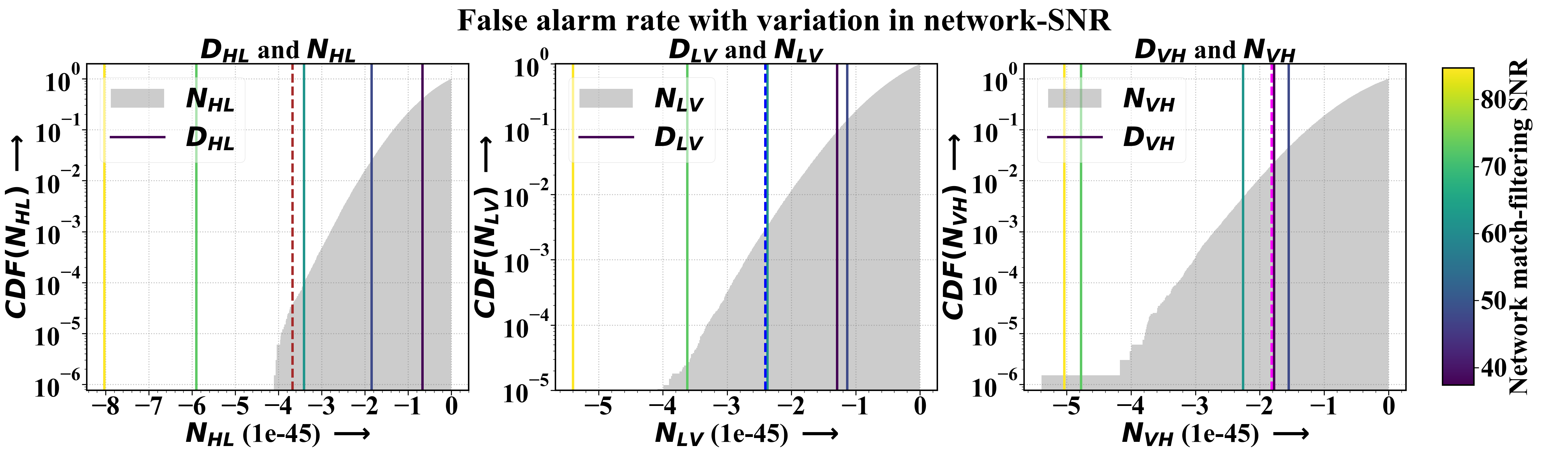}
    \caption{The figure shows the false alarm probabilities for the residual cross-correlation in different detector pair. We have plotted the CDF of the noise cross-correlation distribution (gray) along with different lines showing the residual cross-correlation peak strength by varying the chirp mass of the system. The colour of the lines represent the network matched filter SNR of the WO microlensed signal. Residual cross-correlations from the figure \ref{fig:x} is shown with dashed lines. The values of $a$, $b$, $f_0$ and $k$ are 1.8, 0.4, 30 Hz and 0.004 respectively.}
    \label{fig:FAR}
\end{figure}

On the other hand, we note that the lensing parameters ($b, f_0, k$) show no degeneracy with the source parameters ($m_1, m_2, d_{\rm eff}, \iota, s_{1z}, s_{2z}$), in particular the oscillation parameter $f_0$ shows no correlation with the source masses $m_1$, $m_2$, which control the chirping nature of the signal. We also find that the perturbation parameter $b$ is non-degenerate with the aligned spins $s_{1z}$ and $s_{2z}$ which often introduce beating pattern like envelopes on the waveform. This is evident from the fact that, lensing oscillations ($f_0$) and its strength ($b$) are coming from the properties of the lens whereas the masses and spins of BBH are its intrinsic properties. Note that lensing effects cannot change the GW frequency evolution. The chirping nature of GW frequency evolution is solely dependent on the chirp mass of compact binary. Thus the frequency evolution is controlled by the GW source and frequency-dependent modulations are controlled by the lens properties. Therefore, from a lensed GW, the effect of WO microlensing features can be distinguished from the source imprinted properties, therefore such a correlation between the source and lensing parameters can be mitigated. However, we observe that since $a$ and $d_{\rm source}$ for a lensed event cannot be estimated individually, we combined them in the effective distance parameter $d_{\rm eff} \equiv d_{\rm source}/ a$, which is well constrained in the range around injected $d_{\rm eff}$. Finally, we also note that, there is a correlation between $b$ and $k$, which was also observed in the figure \ref{fig:3}, since the effect of smaller b is equivalent to making b fixed and k larger. 

Here we have considered a prior of $f_0$ up to $1024$ Hz. In appendix \ref{app:a}, we have shown that even with a reduced prior of $f_0$ up to the frequency $f_{\rm isco}$, the results of the joint distributions and the best-fit reconstruction has only been affected marginally for the same noise realization. This indicates that use of a large prior on $f_0$ does not adversely impact the results.

\section{False Alarm Rate for WO microlensing Detection}\label{sec:8}
To understand the significance of a WO microlensing detection, we have assigned a False Alarm Rate (FAR) with the detections. The false alarms are triggered when there is actually no WO microlensing signature present in the data, but \texttt{$\mu$-GLANCE} shows some non-zero cross-correlation signal, caused by the randomness of the noise. We have generated detector noise for Hanford, Livingston, and Virgo with aforementioned noise PSDs for 1 year of observation time. We then performed the cross-correlation between the noise segments of $8$ second length of each with a cross-correlation timescale of $0.5$ second. To calculate the FAR (in units of per year), we obtain the residual cross-correlation signal with varying the source and lens parameters and obtain the number of noise cross-correlation points above the certain residual cross-correlation. 

In figure \ref{fig:FAR}, we have showed the cumulative distribution function (CDF) of the noise cross-correlation. We have plotted different lines to show the residual cross-correlation strength variation with network matched-filtering SNR, by changing the chirp mass of the system. We have also shown the strength of the case we have considered in figure \ref{fig:x}, with dashed lines \footnote{Considering the symmetry of the noise-cross correlation distribution, we have considered the absolute values of the $N_{xx'}$ points and plotted its negative. We also did the same for the residual cross-correlation $x \rightarrow{-|x|}$. This was performed to get the probability of a false alarm probability of a residual cross-correlation directly.}. We observe that with the system of detector frame BBH masses as $(m_1, m_2)=(35, 25)M_{\odot}$ at $1$ Gpc with lensing parameters $a=1.8$, $b=0.4$, $f_0=30$ Hz and $k=0.004$, we obtain a false alarm probability of $4\times 10^{-5}$ with the residual cross-correlation in H-L pair. Similarly for L-V and V-H pair, we obtain a false alarm probability of few times $4 \times 10^{-3}$ and $3 \times 10^{-2}$. Since the noise distribution is generated with $1$ year of simulated noise, the false alarm probabilities directly correspond to the FAR (in yr$^{-1}$ units) corresponding to the detection of the WO microlensing search. 

From figure \ref{fig:FAR} we also note that, with increase in the chirp mass and therefore the SNR of the signal, how the residual cross-correlation corresponds to lower FAR. We observe, above $60$  network-SNR in H-L and above $80 M_{\odot}$ network-SNR in L-V and V-H, the FAR drops below $10^{-3}$ per year, making a robust WO microlensing detection possible. Therefore, the WO microlensing detection is very likely in high-SNR (above 60 network matched-filter SNR at least) systems. We suggest to call an event WO microlensed candidate if the FAR falls below $10^{-3}$ per year in any of the detector pair. In appendix \ref{app:b}, we have showed the effect of different lensing parameters and have estimated the minimum network matched filter SNR required for a low FAR detection ($\leq 10^{-3}$ per year).

\section{Conclusion}\label{sec:9}

In this work, we have demonstrated a new technique \texttt{$\mu$-GLANCE} to search for WO microlensed GW signals from current detectors. In this method, we performed a model-independent search for WO microlensing signatures of the GWs by assuming a unlensed hypothesis and obtaining the best-fit GW strain fitting the data. The best-fit is used to calculate the residual signal in each detector. We cross-correlate the aligned residuals and obtain the residual cross-correlation. By comparing with the noise cross-correlation, we obtained a range of source and lensing parameters which would show up against the noise cross-correlation distribution. We also followed a lensing amplification model based approach for the joint exploration of the source and lensing template parameters together. Notably, with the chosen simple lensing amplification model, we observe that GW source parameters are not degenerate with the lensing parameters. One important caveat of this technique is that, even if the model-dependent source and lensing template parameters joint PE technique can work on single-detector events, the model-independent cross-correlation based-search cannot be performed using a single GW detector data. 

We would like to emphasize the necessity of using an accurate GW waveform model, that is able to capture all the essential features of compact binary (e.g. eccentricity, precession, higher modes). Otherwise, using inaccurate waveform model, we can observe non-zero residual cross-correlations even when there is no WO microlensing present in the GW data. To check whether the waveform model is producing residuals even when there is no WO microlensing, we can put it to a test by checking on a vast different BBH population. If the model is wrong, we may get residuals in many, if not all, of the GW signals and their residual cross-correlations will show up. This is how we can correct waveform model related systematics. However, since the inspiral phase is the longest duration part of the GW signal, and it contributes to most of the signal SNR, any waveform model that can capture the inspiral part well for a coalescing BBH event, can be considered as a good waveform model to search for the lensing signal. However, it is important to point out that though waveform mismodelling may lead to mismodelling of the WO microlensing parameters, but it is unlikely to impact the search WO microlensing search pipeline. This is because strain residual from best-fit will capture both waveform systematics and WO microlensing effects.  

Also, in this analysis, we have not incorporated non-stationarity in the strain noise \citep{Coughlin:2011ck, Mozzon_2020, Mozzon:2021wam}. The non-stationary noise behaviour at one detector is not correlated with the non-stationarity at a different detector, and the duration of a GW signal and the timescale over the which the noise PSD is changing are different. Therefore, we perform cross-correlation on such noise chunks from different detectors at a cross-correlation timescale ($\tau$) much shorter than the variation of the noise-PSD and much longer than a glitch. In this way, we can mitigate the noise fluctuations from generating significant deviations from zero residual cross-correlation \footnote{We also apply bandpass filter to remove very low frequencies where noise-fluctuation is very high and very high frequencies where the jitters are very faster than the signal oscillations} .

The current analytic lensing model we use in this analysis, captures most of the features present in WO microlensing amplification in general. However, we do not claim universality to this amplification template. It shows to be working nicely with the two test cases we used in \ref{sec:3}. It may fail badly for some lens models present in the universe. However this would not impact the residual cross-correlation since, obtaining the best-fit strain at a detector and from there obtaining the residual and finally performing the residual cross-correlation  -- each step is done in a model independent approach. Therefore, when the amplification template fails, although hinders the joint parameter space exploration side, does not affect the WO microlensing detection side, which is a completely data-driven model-agnostic approach for a robust technique to detect WO microlensed GWs. However, we have not associated any of the WO microlensing parameters $a$, $b$, $f_0$, $k$ to physical space parameters in the lens system, like the offset parameter $y$, the mass of the lens $M_{l}$ or its density profile $\rho(\vec{r})$. This will be performed in a future work.

With the gradual advancement of the ground-based GW detector sensitivities, and the inclusion of the next-generation like Cosmic Explorer (CE) \citep{Evans:2021gyd}, Einstein Telescope (ET) \citep{Maggiore:2019uih} or Laser Interferometer Space Antenna (LISA) \citep{LISA:2022yao}, the WO microlensing signatures in the cross-correlation is expected to be predominant over noise cross-correlation distribution. This will allow WO microlensed GW signals to be detected at an unprecedented precision. The WO microlensing amplification can lead to inferring the intrinsic properties of the lensing object. Therefore, under the wave-optics regime, the WO microlensing aspects would allow us to probe the dark matter haloes and sub-haloes, which are otherwise invisible to the electromagnetic signal. In this regard, \texttt{$\mu$-GLANCE} will play a pivotal role in both the detection and precise characterization of WO microlensing signatures, significantly enhancing our ability to identify and analyse these phenomena with greater efficiency.
\section{Acknowledgement}

The authors are thankful to Sourabh Magare for reviewing the manuscript
during the LSC Publications and Presentations procedure and providing useful comments. The authors express their gratitude to the \texttt{⟨data|theory⟩ Universe-Lab} group-members for useful suggestions. This work is part of the \texttt{⟨data|theory⟩ Universe-Lab}, supported by TIFR and the Department of Atomic Energy, Government of India. The authors express gratitude to the computer cluster of \texttt{⟨data|theory⟩ Universe-Lab} for computing resources used in this analysis. We thank the LIGO-Virgo-KAGRA Scientific Collaboration for providing noise curves. LIGO, funded by the U.S. National Science Foundation (NSF), and Virgo, supported by the French CNRS, Italian INFN, and Dutch Nikhef, along with contributions from Polish and Hungarian institutes. The research leverages data and software from the Gravitational Wave Open Science Center, a service provided by LIGO Laboratory, the LIGO Scientific Collaboration, Virgo Collaboration, and KAGRA. Advanced LIGO's construction and operation receive support from STFC of the UK, Max-Planck Society (MPS), and the State of Niedersachsen/Germany, with additional backing from the Australian Research Council. Virgo, affiliated with the European Gravitational Observatory (EGO), secures funding through contributions from various European institutions. Meanwhile, KAGRA's construction and operation are funded by MEXT, JSPS, NRF, MSIT, AS, and MoST. This material is based upon work supported by NSF’s LIGO Laboratory which is a major facility fully funded by the National Science Foundation. We acknowledge the use of the following python packages in this work: NUMPY \citep{harris2020array}, SCIPY \citep{2020SciPy-NMeth}, MATPLOTLIB \citep{Hunter:2007}, ASTROPY\citep{The_Astropy_Collaboration_2022}, PYCBC \citep{alex_nitz_2024_10473621}, GWPY \citep{gwpy}, LALSUITE \citep{lalsuite}, EMCEE \citep{Foreman_Mackey_2013} and CORNER \citep{corner}, BILBY \citep{Ashton_2019}.

%





\bibliography{biblio}{}
\bibliographystyle{aasjournal}



\appendix

\section{Appendix: Inclusion of amplification template phase term \texorpdfstring{$\phi$}{phi}
in the source and lensing effects joint Parameter Estimation}\label{app:0}
\begin{figure}
    \centering
    \includegraphics[width=0.8\linewidth]{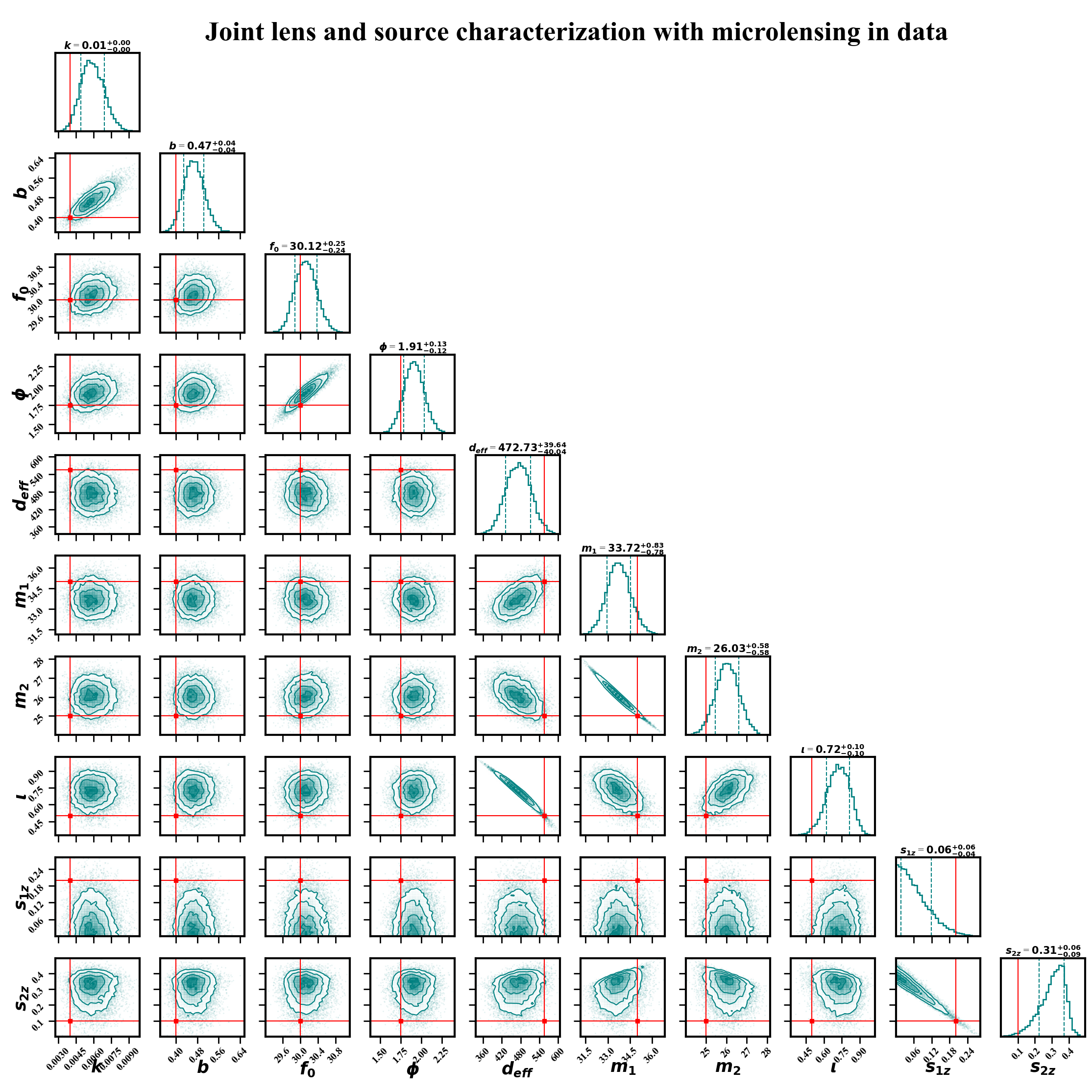}
    \caption{The figure shows the estimation of the amplification phase together with the WO microlensing parameters and source parameters as a part of joint parameter space exploration.}
    \label{fig:pe_phase}
\end{figure}
Here we have included the phase of the amplification $\phi$ in the joint parameter exploration. The figure \ref{fig:pe_phase} shows that, with the inclusion of the phase of the amplification, which was not considered in the analysis, for obvious reasons, does not affect the rest of the source and lens parameters. This shows that the phase term $\phi$ has no correlation with any of the other source and lens parameters and our results without the phase are absolutely fine to consider.

\section{Appendix: Effect of Different Choices of \texorpdfstring{$f_0$}{f0} Priors on Parameter Estimation}\label{app:a}

To understand whether we have effects on the posteriors caused by the choice of the priors, especially for the case when there was no WO microlensing  injected in the data. We study the effect of the choice of the parameter $f_0$, for which we have two different priors: (i) $f_0 = [20, 1024]$ Hz and (ii) $f_0 = [20, f_{\rm isco}]$ Hz, where $f_{\rm isco}$ is the innermost stable circular orbit frequency of the binary black holes and $1024 Hz$ is the final frequency of the the frequency-series data. 

\begin{figure}
    \centering
    \includegraphics[width=0.8\linewidth]{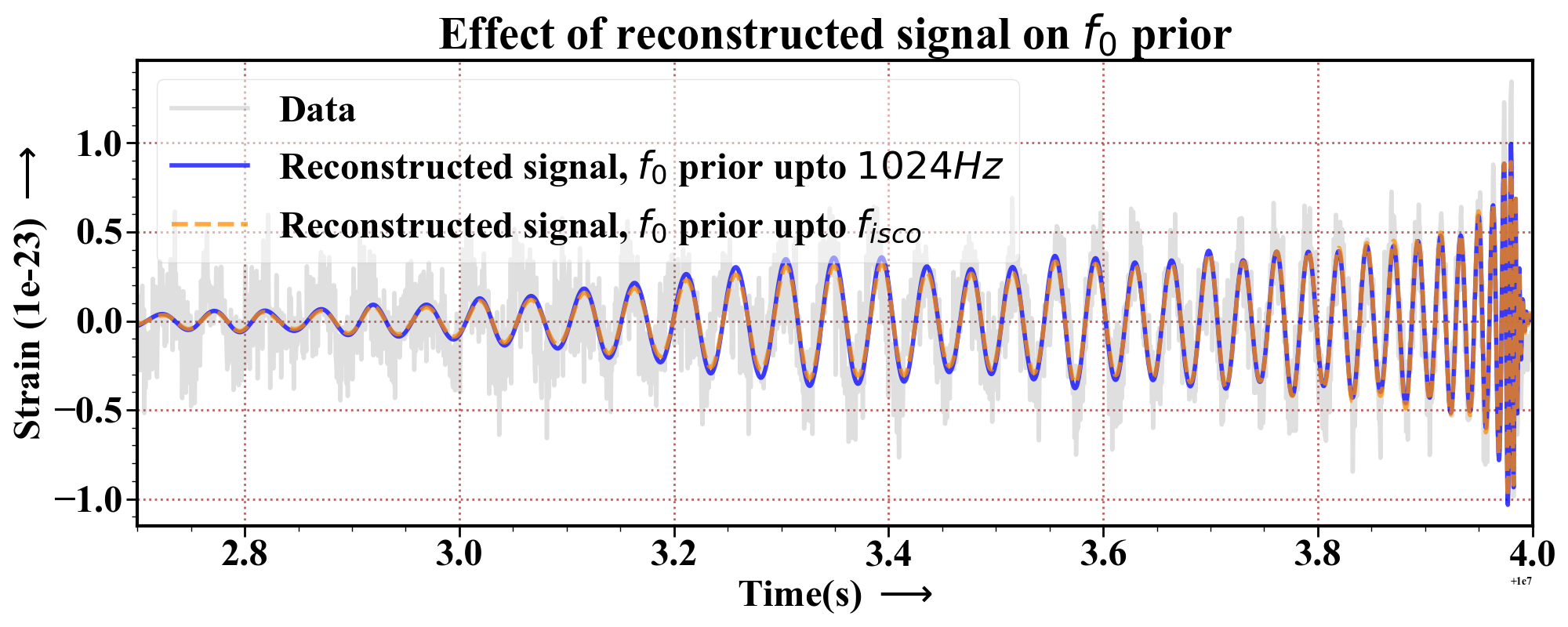}
    \caption{Comparison between two different priors on $f_0$ on the reconstructed signal when the actual data had no WO microlensing features. We observe that the reconstructions do not show any significant differences in the WO microlensing features and therefore are very similar to each other.}
    \label{fig:n}
\end{figure}

\begin{figure}
    \centering
    \includegraphics[width=0.8\linewidth]{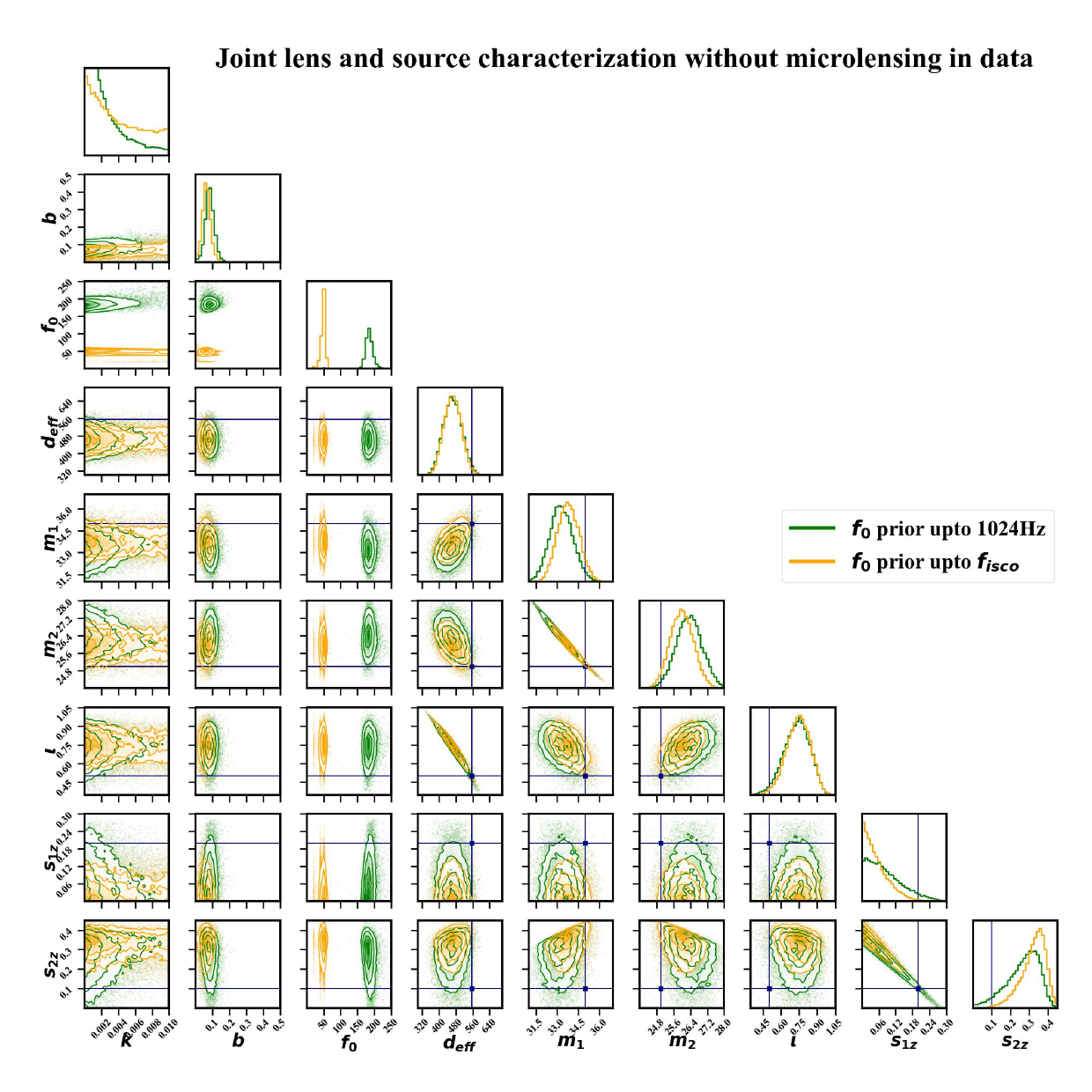}
    \caption{The corner plot shows the joint distribution of source and WO microlensing template parameters obtained for the two choices of priors mentioned, when the signal was not WO microlensed. The parameter space occupied by these two cases, although very similar for the source parameters, shows slight disagreement for the lensing parameters. }
    \label{fig:no_ML}
\end{figure}

\begin{figure}
    \centering
    \includegraphics[width=0.8\linewidth]{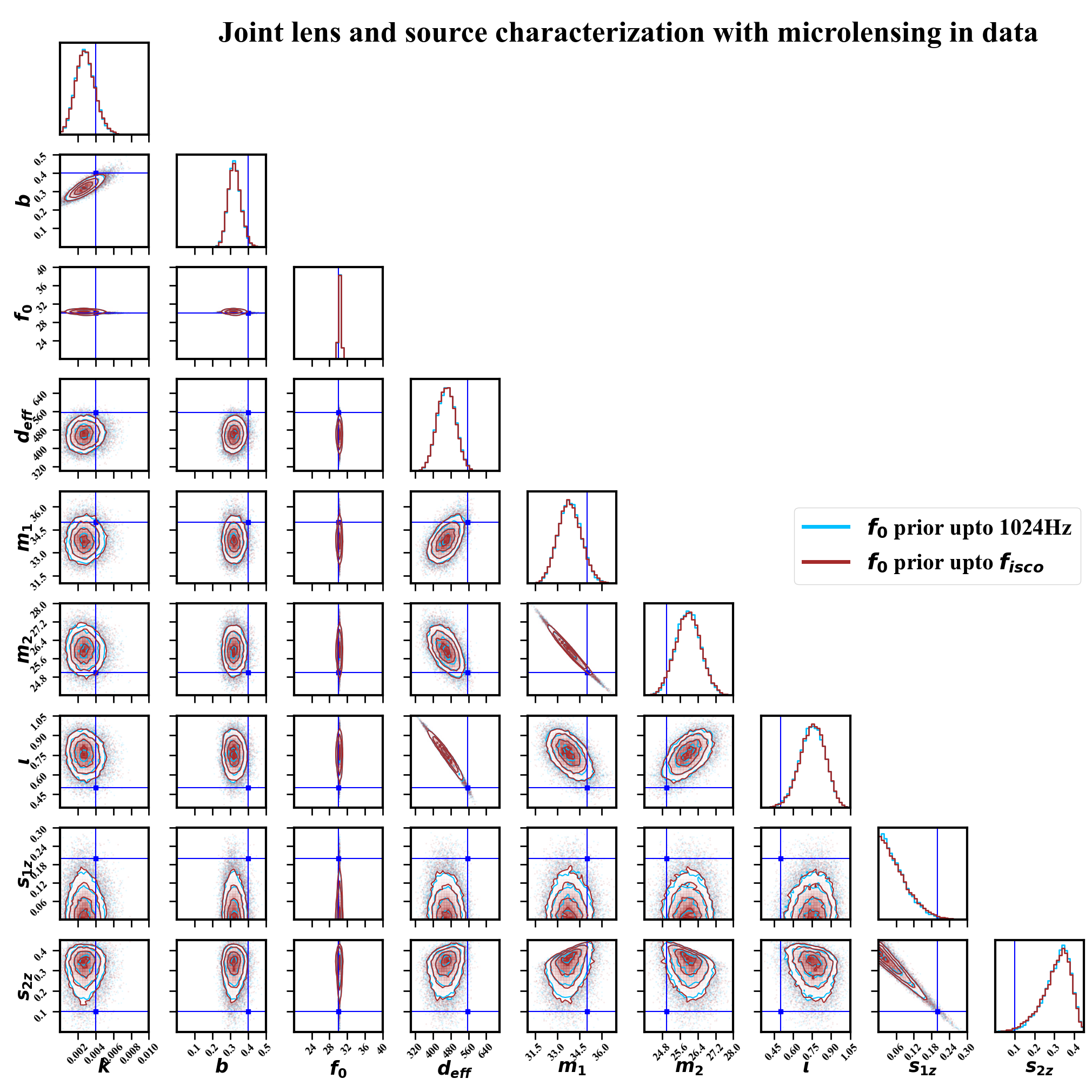}
    \caption{The corner plot shows the joint distribution of source and WO microlensing parameters obtained for the two choices of priors mentioned, when the signal was WO microlensed. We observe that the choice of prior does not shift any of source or lensing parameters, if not shifted marginally.}
    \label{fig:ML}
\end{figure}

We have plotted the frequency-domain data and the waveform model is constructed with the medians of the posteriors. To compare with we have shown in figure \ref{fig:n} that, we do not observe any direct effect of the choice of $f_0$ prior, since the medians of the WO microlensing perturbation parameter $b$ was very close to being zero. The medians of $b$ for both the cases with $f_0$ prior up to $1024$ Hz and $f_{\rm isco}$ are 0.0877 and 0.0764 respectively, therefore not showing any WO microlensing features prominently. We also obtained the joint distribution of the WO microlensing template parameters and source parameters when the signal was not WO microlensed, with those two different $f_0$ priors. The comparison of the choice of such different prior choices is shown in figure \ref{fig:no_ML}. Similarly, we obtained the joint distribution of the WO microlensing template parameters and source parameters when the signal was actually WO microlensed, with those two different $f_0$ priors. The comparison of the choice of such different prior choices is shown in figure \ref{fig:ML}.

\section{Appendix: False Alarm Rates Depending on Source and Lens Parameters}\label{app:b}

In figure \ref{fig:FAR}, we obtained the FAR for a particular combination of lens parameters and varied the strength of the signal by varying the chirp mass of the compact binary. Here we present a few more combinations of the lensing parameters, it shows the confidence at which we are going to detect them. 

We fixed the oscillation parameter $f_0$ since from figure \ref{fig:3} we observed that the residual cross-correlation signal has no direct dependency on the value of $f_0$. Therefore, we only varied $a$, $b$, and $k$ to show at what FAR we can detect WO microlensing events.

\begin{figure}
    \centering
    \includegraphics[width=0.7\linewidth]{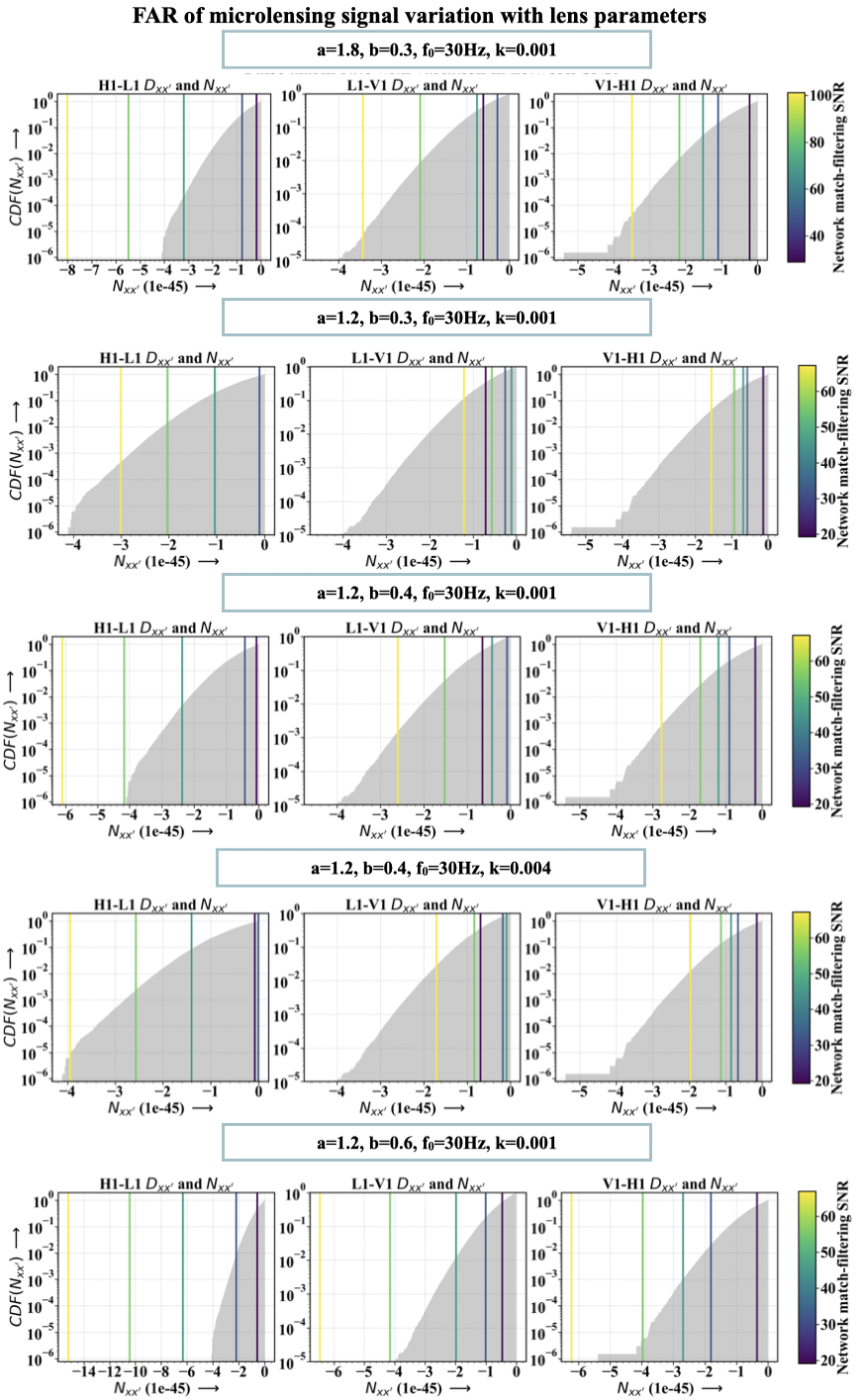}
    \caption{The figure shows the variation of the FAR of WO microlensing signal detection for a couple of more variations on lensing amplification parameters $a$, $b$, and $k$. This helps us understand for a wide variety of lensing amplification effects, how confidently we are going to claim a lensing detection for a given network SNR.}
    \label{fig:b1}
\end{figure}

In figure \ref{fig:b1}, we varied the network matched-filtering SNR and observe the dependency of FAR of the WO microlensing detection, depending on the values of $a$, $b$, and $k$. We show on the first pane for $a=1.8$ at $b=0.3$ and $k=0.001$, it shows that close to network SNR of 70, we can have a FAR on at least one detector pair below $10^{-3} per year$. For the case considering $a=1.2$, $b=0.3$ and $k=0.001$, the minimum network SNR is still at around 65 to detect a WO microlensed signal. This is because the effect of the strong lensing magnification $a$ has been absorbed in the luminosity distance while performing the parameter estimation for the best-fit. For the case with $a=1.2$, $b=0.4$ and $k=0.001$, we observe that the minimum network SNR is close to 45, where for the same $a$ and $b$ values, with $k=0.004$, it is moved back to 55. Finally, with $a=1.2$, $b=0.6$ and $k=0.001$, the minimum network SNR required is about 35. 

From this study, we can conclude that WO microlensing is hard to detect with current generation GW detectors unless the SNR of the source is very high. With the inclusion of more detectors and better detector sensitivities such as LIGO-India \citep{LIGO_India}, A$^\#$ \citep{A_hash}, CE\citep{Hall:2020dps} or ET\citep{Punturo:2010zz} observatories, detection of WO microlensed events will be more feasible. The availability of more detectors with better detector sensitivity will help in obtaining cross-correlation signals between more pairs of detectors ($N_{\rm det}(N_{\rm det} -1)/2$, where $N_{\rm det}$ is the number of detectors) and will also help in reducing the FAR. As a result, it enables for a robust detection of the WO microlensing signatures from GW signals.

\end{document}